\documentclass[%
 reprint,
 amsmath,amssymb,
 aps,
floatfix,
]{revtex4-1}

\usepackage{graphicx}
\usepackage{dcolumn}
\usepackage{bm}
\usepackage{color}

\usepackage{xparse}
\usepackage{tikz}
\usetikzlibrary{matrix,backgrounds}
\pgfdeclarelayer{myback}
\pgfsetlayers{myback,background,main}

\tikzset{mycolor/.style = {line width=1bp,color=#1}}%
\tikzset{myfillcolor/.style = {draw,fill=#1}}%

\NewDocumentCommand{\highlight}{O{blue!40} m m}{%
\draw[mycolor=#1] (#2.north west)rectangle (#3.south east);
}

\NewDocumentCommand{\fhighlight}{O{blue!40} m m}{%
\draw[myfillcolor=#1] (#2.north west)rectangle (#3.south east);
}

\begin{document}

\preprint{APS/123-QED}

\title{Dynamical properties of hierarchical networks of Van Der Pol oscillators}

\author{Daniel Monsivais$^1$}
\author{Kunal Bhattacharya$^{1,2}$}
\author{Rafael A. Barrio$^3$}
\author{Philip K. Maini$^4$}
\author{Kimmo K. Kaski$^{1,4,5}$}


\affiliation{$^1$Department of Computer Science, Aalto University School of Science, 00076, Finland}
\affiliation{$^2$Department of Industrial Engineering and Management, Aalto University School of Science, 00076, Finland}
\affiliation{$^3$Instituto de  F\'isica, U.N.A.M., 01000, Ap. Postal 101000, M\'exico D.F., M\'exico}
\affiliation{$^4$Wolfson Centre for Mathematical Biology, Mathematical Institute, Oxford University, Oxford, UK}
\affiliation{$^5$The Alan Turing Institute, 96 Euston Rd, Kings Cross, London NW1 2DB, UK}

\date{\today}

\begin{abstract}
Oscillator networks found in social and biological systems are characterized by the presence of wide ranges of coupling strengths and complex organization. Yet robustness and synchronization of oscillations are found to emerge on macro-scales that eventually become key to the functioning of these systems. In order to model this kind of dynamics 
observed, for example, in systems of circadian oscillators, we study networks of Van der Pol oscillators that are connected with hierarchical couplings. 
For each isolated oscillator we assume the same fundamental frequency. Using numerical simulations, we show that the coupled system goes to a phase-locked state, with both phase and frequency being the same for every oscillator at each level of the hierarchy. The observed frequency at each level of the hierarchy changes, reaching an asymptotic lowest value at the uppermost level. Notably, the asymptotic frequency can be tuned to any value below the fundamental frequency of an uncoupled Van der Pol oscillator. We compare the numerical results with those of an approximate analytic solution and find them to be in qualitative agreement. 

\end{abstract}

\maketitle

\section{INTRODUCTION}

Oscillations are commonly observed in various physical, chemical and biological systems~\cite{osipov2007synchronization}. The reason is simple; when there are two fields that compete with each other in such a way that the rate of change of one of the fields is proportional to the strength of the other field opposing the first one, oscillatory behavior follows. In systems where one can find many constituents that are inherently oscillatory and are coupled to each other for the regulation of multiple functions, complex dynamical behavior emerge~\cite{boccaletti2002synchronization,abrams2004chimera,strogatz2000kuramoto}. 

A system with couplings between the oscillatory elements can be visualised as a network, the properties of which 
dictate the regulation of behavior at increasing levels of complexity~\cite{arenas2008synchronization}. For instance, if one considers an arbitrary network of simple harmonic oscillators coupled with linear interactions of the same strength, then the fundamental frequency of oscillation 
of the system decreases with the number of oscillators $N$, as $1/\sqrt{N}$. This means that the properties of the system depend crucially on the number of components, a feature that is not desirable, particularly in biological processes~\cite{goldbeter1997biochemical,glass2001synchronization}.

This is the case, for example, with the biological clock of a living organism. It is well known that individual cells in the suprachiasmatic nucleus (SCN) of the mammalian brain are circadian oscillators, each having the correct 24 hour period of the biological clock~\cite{forger2017biological,liu1997cellular}. In the SCN there are thousands of cells (or circadian oscillators) that are coupled to each other, and these interactions allow the regulation of many circadian rhythms, including the circasemidian or semicircadian rhythm of 12 hours, through complex mechanisms~\cite{garcia2004modeling}. It is understood that one of the most fundamental and useful properties of these oscillators, namely the frequency of 1/day, does not vary with the number of coupled oscillators in the network, which leads to the conclusion that the network structure of this system cannot be simple. This, in turn, emphasises the importance of developing descriptive models for oscillator networks whose frequency does not change for a large number of coupled components.

In a previous study some of us~\cite{barrio1997hierarchically}
proposed a model of networked oscillators that is able to stabilise the fundamental frequency at any desired value by considering a hierarchy of coupling strengths between the oscillator nodes. There the mean field calculations showed that one could obtain circadian oscillations from ultradian ones with periods of the order of milliseconds. 
These results could be relevant in a different context. 
For example, the daily rhythm of 
individuals in a network of social interactions should not change with the number of individuals. Undoubtedly, the circadian behavior of a society follows not only the sun, but also the social pressures and interactions between individuals~\cite{grandin2006social}. In particular, the natural circasemidian or semicircadian rhythms of humans have to adjust to a circadian pace due to the social pressure of the surrounding society. This has been observed in recent studies, see for example,~\cite{monsivais2017seasonal,monsivais2017tracking}, in which 
the interplay between social and environmental factors is 
reflected in the circadian rhythms of human communication. Also, recently  
several experiments with humans and other species of animals have demonstrated the influence of social interactions on the circadian rhythms of individuals and groups of individuals~\cite{mistlberger2004social,bloch2013socially,fuchikawa2016potent}.

In this paper we propose a model of coupled Van der Pol oscillators in a family of networks that exhibits not only the stabilisation of the frequency, but also synchronization and phase locking, which are properties that are indispensable in modelling many biological processes. The main aim of our model is to show a way in which a system of coupled 
oscillators could stabilize its fundamental frequency, such that this frequency does not depend on the number of oscillators when their number is large. There has been a number of studies of the properties of Van der Pol oscillators, coupled in various ways~\cite{pastor1993ordered,ulonska2016chimera,dorfler2014synchronization}.
In these works, the hierarchy of the networks is introduced via topological features, reflected in the adjacency matrix~\cite{perlikowski2010discontinuous,hizanidis2015chimera,bera2016imperfect,krishnagopal2017synchronization,rakshit2018synchronization}. On the contrary, in this manuscript, the hierachical nature of the system is imposed assigning different couplings strengths (weights) to the links joining different parts of the network. To the best of our knowledge, our work is the first that uses this approach. In addition, our model aims to give insight into the properties of a complex network of hierarchically coupled Van der Pol oscillators in various regular topological structures. 

This paper is organised  as follows. After the Introduction (section I)  we describe our hierarchical network model (section II). Then we present the numerical solutions for various hierarchical network topologies (section III). We also develop a mean-field-like approximate analytical solution for all these hierarchical network models (section IV). Finally in section V we present our concluding remarks.

\section{THE MODEL}

Our model is based on a previous one formulated by some of the authors of this study~\cite{barrio1997hierarchically}, in which a hierarchically connected group of $N$ oscillators is shown to stabilize the collective frequency of oscillation, while avoiding the universal feature that the frequency decays like 
$\approx 1/\sqrt{N}$. This is a peculiar phenomenon, since a set of linear (or non-linear) oscillators with constant interactions between the nodes of an arbitrary network exhibits a fundamental frequency that follows the dependence 
$\omega(N)=\omega(0) \sqrt{1-r(N-1)}$, where $r$ is the strength of the elastic interactions and $\omega(0)$ is the fundamental frequency of a single oscillator. Note that if $r \geq (N-1)^{-1}$ then the system becomes unstable.

\begin{figure}
    \centering
    \includegraphics[width=0.45\textwidth]{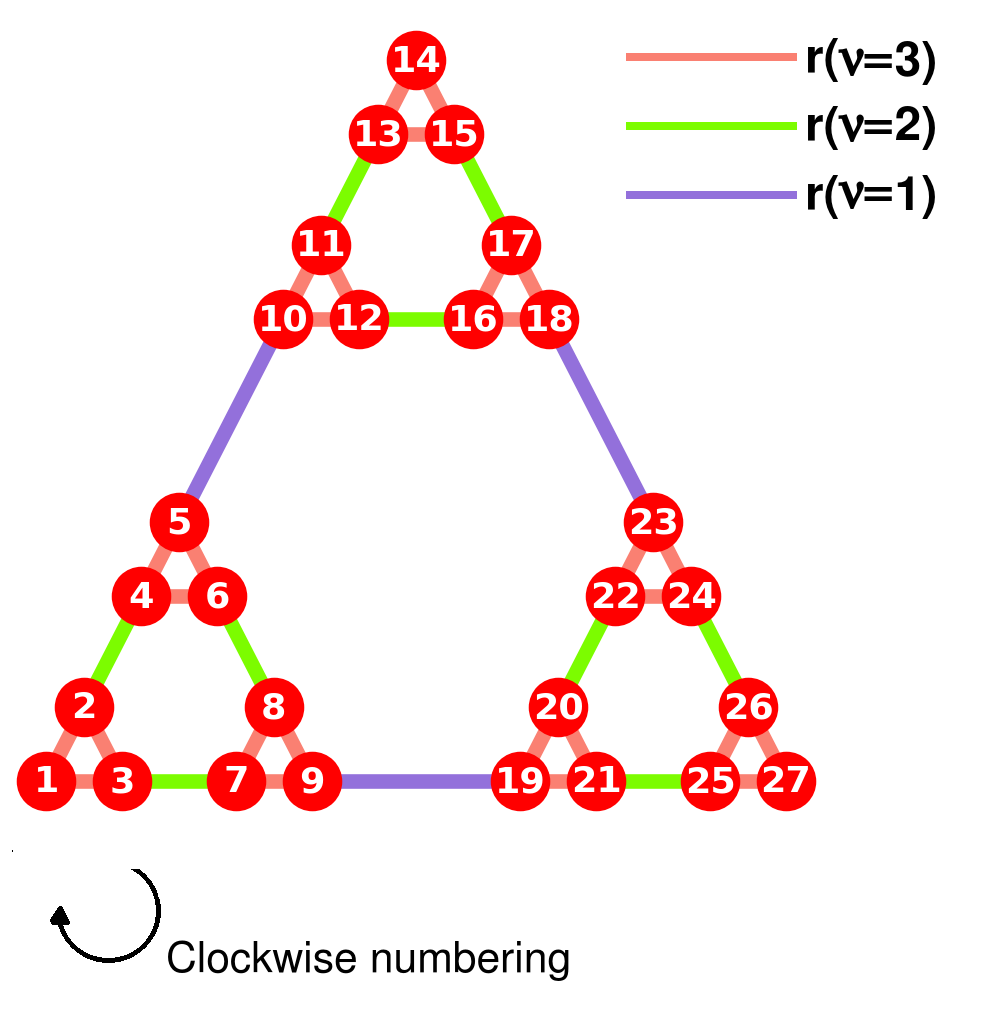}
    \includegraphics[width=0.49\textwidth]{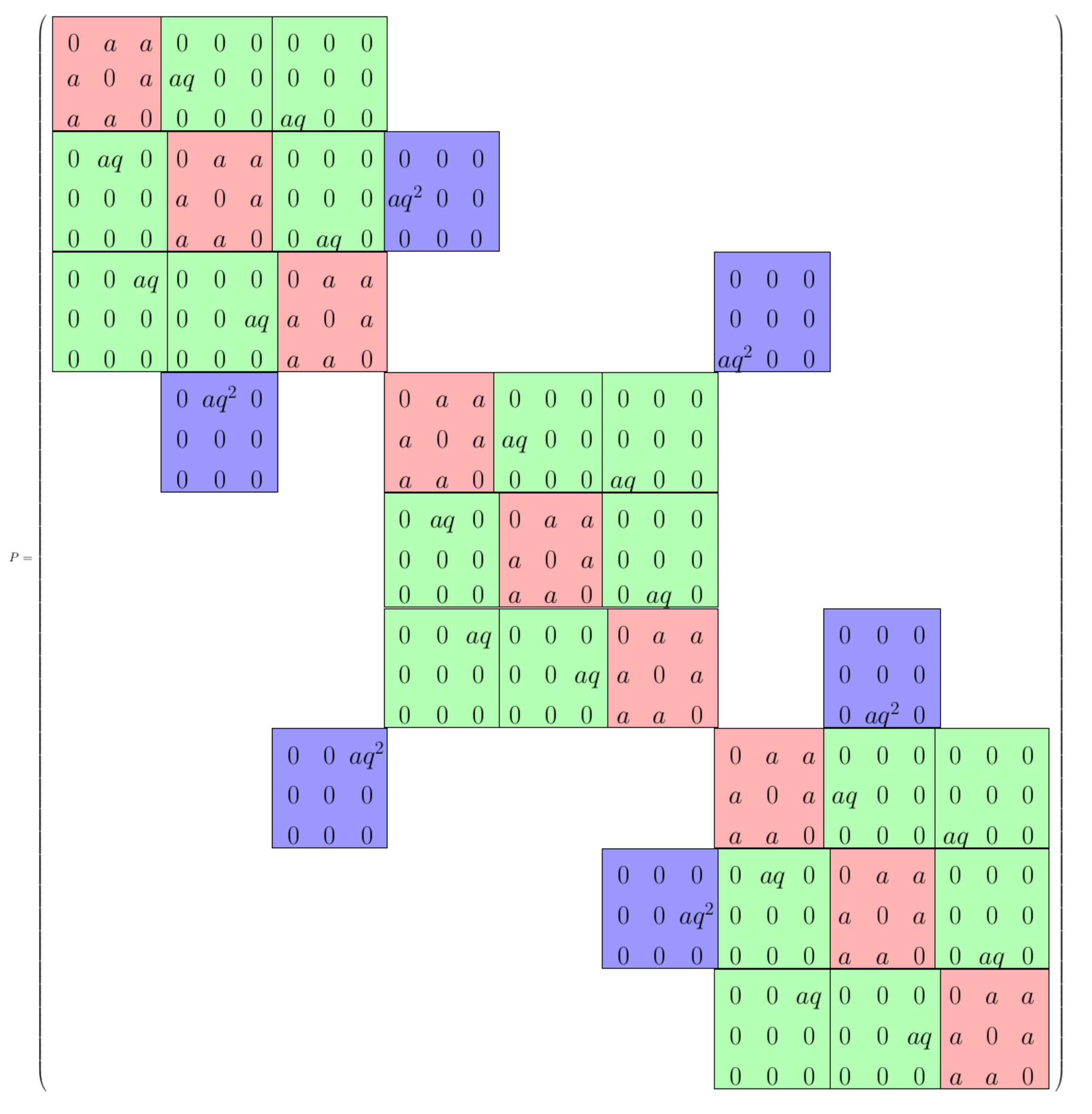}
    \caption{(Top) Recursively generated fractal-like network  with $L=3$  levels and coordination number $K=3$. The network ($\mathbb{S}(3,3)$) has triangular topology and it contains triads of fundamental oscillators as the basic units $K=3$-cliques that are coupled up to the level $L=3$. Links joining oscillators inside each basic unit or clique are colored in pink, and links joining these basic triangular units are colored in green, thus forming yet larger triangular units of basic triangular units, while links joining these larger units of basic units are colored in purple forming triangular units of triangular units of basic triangular units. (Bottom) Adjacency matrix representing the network ($\mathbb{S}(3,3)$) shown in the top panel. For the number of levels $L=3$ the network has $ 3^L = 27$ oscillators. Colored blocks encircle the links between each basic triplet of oscillators and the other elements in the network. The color code is the same as above 
    with pink, green and purple representing links within each unit, between units, and between units of units, respectively.}
    \label{fig:1a}
\end{figure}

\begin{figure}[ht!]
    \centering
    \includegraphics[trim= 0cm 0.25cm 0cm 0.25cm , clip,width=0.4\textwidth]{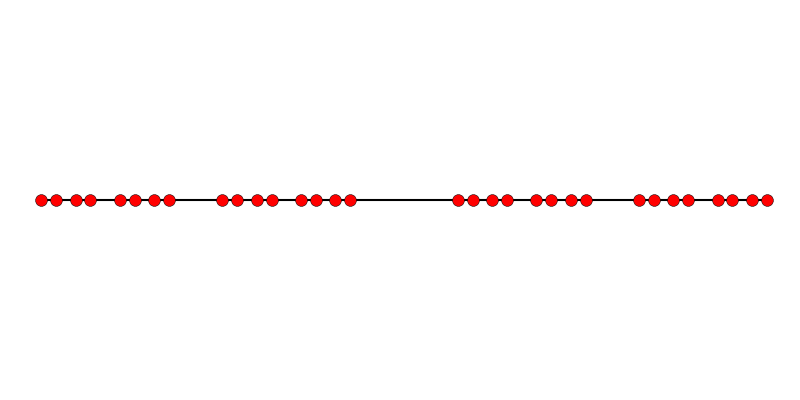}
    \includegraphics[width=0.2\textwidth]{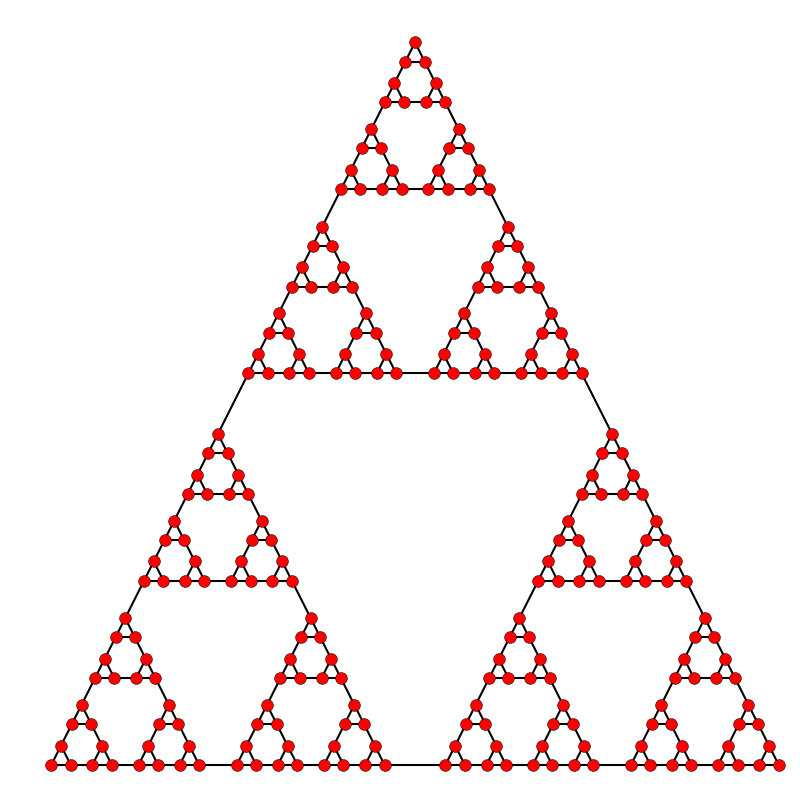}
    \includegraphics[width=0.2\textwidth]{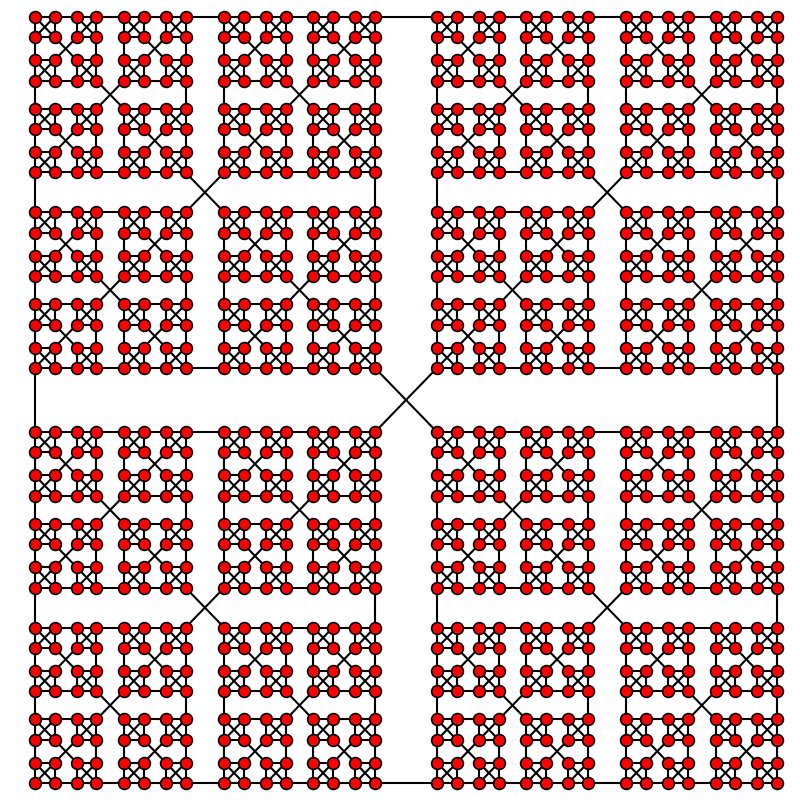}
    \includegraphics[width=0.2\textwidth]{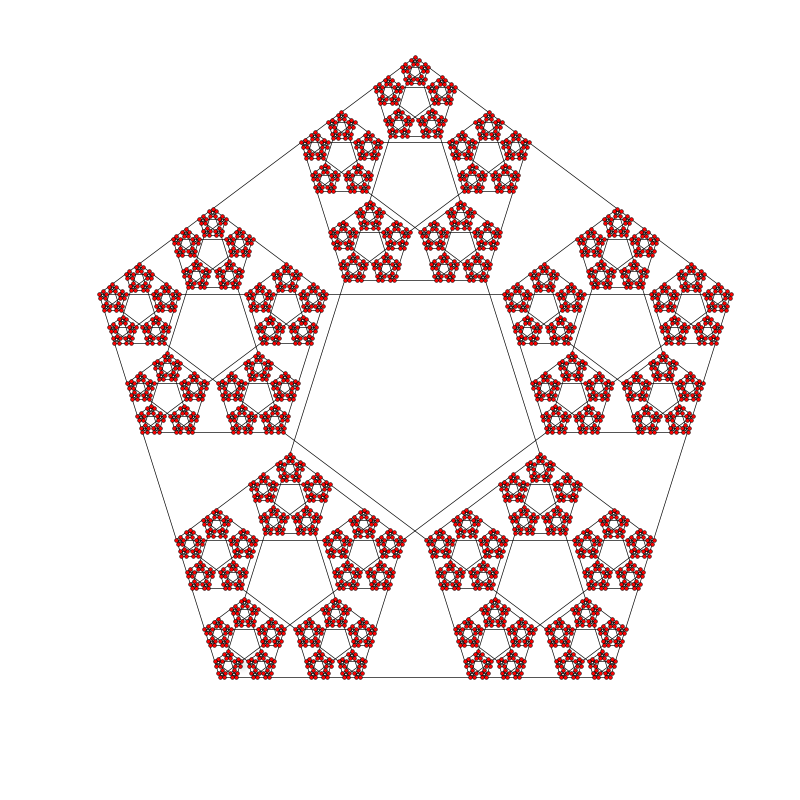}
    \includegraphics[width=0.2\textwidth]{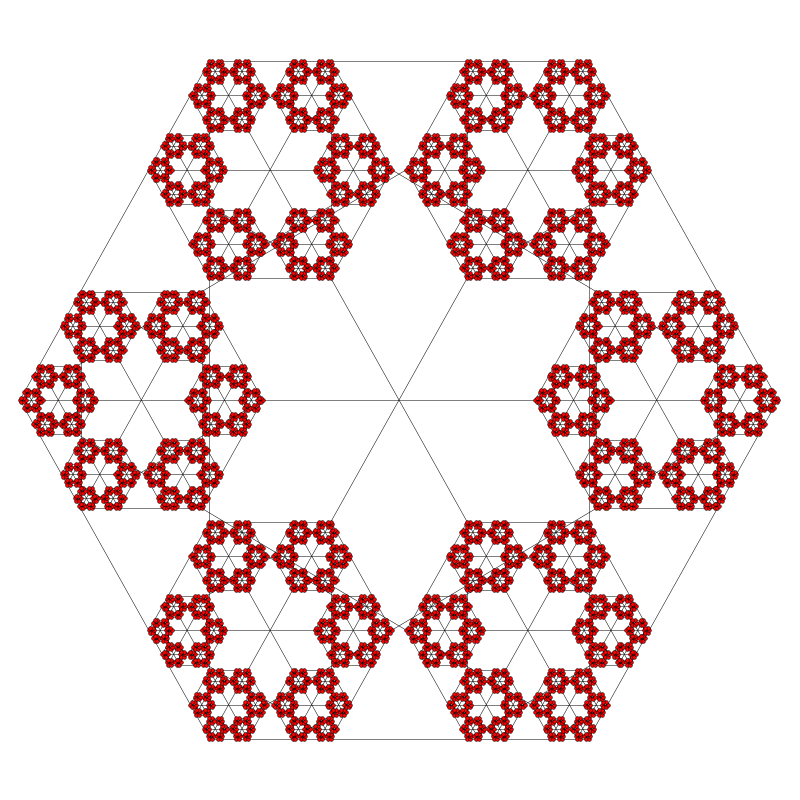}
    \caption{Examples of five different topology Sierpinski graphs $\mathbb{S}(L,K)$, for $L=5$ levels and $K = 2, 3, 4, 5, 6$; (top) $K=2$ and $2^L=32$ nodes; (middle-left) $K=3$ and $3^L=243$ nodes; (middle-right) $K=4$ and $4^L=1024$ nodes; (bottom-left) $K=5$ and $5^L=3125$ nodes; (bottom-right) $K=6$ and $6^L=7776$ nodes.}
    \label{fig:2}
\end{figure}

In our model we assume the oscillator nodes to be of non-linear Van der Pol type~\cite{van1927vii}, described by the following equation~\footnote{The Van der Pol equation is a special case of the Rayleigh Differential equation} 

\begin{equation}\label{eq:1}
    \ddot{x} - v(1-x^2)\dot{x} +\omega_0^2 x = 0 
\end{equation}
where $v > 0$ is the bifurcation parameter and $\omega_0$ the frequency parameter.
For a system of $N$ identical oscillators coupled with elastic connections $r_{ij}$ (in units of the frequency $\omega_0$), the dynamics can be expressed by the following set of equations,

\begin{subequations}
\begin{align}
    \dot{x}_i=&\omega_0 y_i, \label{eq:2a}\\
    \dot{y}_i=&-\omega_0 ( x_i - \sum_{j \in n(i)} r_{ij} x_j ) + v(1-x_i^2)y_i\label{eq:2b},
\end{align}
\end{subequations}
where $n(i)$ is the set of neighbors of an oscillator $i$, placed in a node of a regular graph with fractal-like topology of $L$ levels and coordination number $K$ (see the top panel in Fig.~\ref{fig:1a}, for $L=3$ and $K=3$). The fractal-like layout of the network is used to determine the strength of the interactions $r_{ij}$ between neighboring oscillators, assigning them variable weights depending on the relative location of the linked nodes in the network. The network consists of many units, each one with $K$ nodes connected in all-to-all fashion ($K$-cliques). These units are, in turn, connected with other units following the same recipe, forming $K$-cliques of units. This recursive process is repeated $L$ times, i.e. the number of levels, to generate a fractal-like connected network, denoted by $\mathbb{S}(L,K)$, and being of size $K^L$. Here we follow the nomenclature of the well known finite Sierpinski graphs~\cite{klavvzar1997graphs}. In Fig.~\ref{fig:2} we show five topologically different $\mathbb{S}(L,K)$ graphs for coordination numbers, $K=$ 2, 3, 4, 5 and 6, each one of them with $L=6$ hierarchical levels or interaction strengths between neighboring oscillators.  

The definition of the hierarchical link strength $r_{ij}$ follows a 
recursive procedure such that the strength of interaction depends on the level $\nu=L, L-1,..., 1$, at which 
the link is located, and is given by the following relation 

\begin{align}
r_{ij}(\nu)&= a q^{L-\nu} ,\,\,\,\,\,\, 0 \leq q \leq 1
\end{align}
%
%
%
where $a$ is the strength at the deepest level $\nu=L$ (while $\nu=1$ represents the shallowest level) and $q$ is a decay parameter of the strength, and independent of $\nu$. Hence, initially all the links connecting nodes inside a single unit ($K$-clique) have strength $r_{ij}(\nu=L) = a$. 
These links belong to the deepest level $\nu=L$ of the network,  thus are the strongest. Next, the links connecting different cliques, i.e. forming $K$-cliques of $K$-cliques, have now diminished strength, $r_{ij}(\nu=L-1) = a q = q r_{ij}(L)$ 
and constitute 
the second deepest level $\nu=L-1$ of the hierarchical network. 
Following this idea, links connecting the cliques of the cliques of the cliques, keep having diminishing strengths as follows 
$r_{ij}(\nu=L-2)= a q^{L-(L-2)}= a q^2$ 
constituting the third deepest level. 
This process continues until the links connecting the largest possible units (the cliques of the cliques of ... of the cliques), constituting the most shallow 
level $\nu=1$, and having the weakest interaction strength, i.e. 
$r_{ij}(\nu=1) = a q^{L-1}$. In Fig.\ref{fig:1a} the connection strengths of a hierarchical network with $K=3$ (i.e. \textsl{triangles as basic units}) are shown, where the strength of connections located at different depth levels $\nu=$ 3, 2, 1 are colored in pink, green and violet, respectively.

To illustrate the structural properties of our model system, let us consider the example of Fig.~\ref{fig:1a} for a network $\mathbb{S}(L=3,K=3)$. Here the three nodes at the three corners of the biggest triangle in Fig.~\ref{fig:1a}(top) have one link less than the rest of the nodes in the network, i.e. their coordination number is two. Similarly for the general topology of the finite network, $\mathbb{S}(L,K)$, having $L$ hierarchical levels and coordination number $K$, there are $K$ corner points or nodes with coordination number $(K-1)$ and a total of $(K^L-K)$ nodes with coordination number $K$. Thus the role of these $K$ corner nodes can be assumed small for large $L$ and they can be used to introduce external force or perturbation into the system. If we number the oscillators in a clockwise manner within each triangle we obtain the weighted adjacency matrix shown in Fig.~\ref{fig:1a}(bottom).

It is obvious that there is no analytic solution to the dynamics of the hierarchically coupled network model presented in Eqs.~\ref{eq:2a}--\ref{eq:2b}, due to its non-linearity and structural complexity. Thus to describe the evolution of the system in time, these equations of motion need to be integrated numerically. To do this we have chosen the fourth order Runge-Kutta method (RK4), 
as it has been found to be well suited for integrating systems of coupled oscillators.

\section{NUMERICAL SOLUTION}
In this network of hierarchically connected oscillators synchronization may occur at different regions and scales as a consequence of interactions between the oscillators. 
We are interested in finding the behavior of the fundamental frequency of the oscillating units in different regions of the network, since these modes are the ones corresponding to the long wavelength excitations and are the modes that persist longer in the network. In order to track and describe such a process, we follow a coarse-grained analysis.

The process starts in a base network $\mathbb{S}(L,K)$ with $L$ levels and link weights as described before. In this base network, the elements forming the basic $K$-cliques (connected by the strongest connections) 
are  considered to oscillate collectively as a single unit, with frequency $\omega(\nu=L\!-\!1)$ which is in general different from the intrinsic frequency of the individual oscillators $\omega(\nu=L)\equiv w_0$. 
Each new unit has an associated output $s^{(\nu)}\equiv s^{(L-1)}$ defined as the sum of the amplitudes of the internal oscillators in the clique forming the unit. In addition, the set of oscillating units can be visualized as a 
network, with 
topology similar to that of the base network but having 
one level less, and considered as a 
Sierpinski graph $\mathbb{S}(L-1,K)$ coarse-grained from the base network. Following this recursive coarse-graining analysis procedure, in the next step a new graph is constructed from the previous $\mathbb{S}(L-1,K)$ graph 
by considering each one of its cliques of nodes as a new single unit with output $s^{(L\!-\!2)}$,  and then this set of new oscillating units is visualized as a network of Sierpinski graph topology $\mathbb{S}(L-2,K)$ with one level less than the former (i.e. two levels less than the original). 
This procedure is repeated until the last step of the coarse-graining process is reached, i.e. a $\mathbb{S}(0,K)$ graph (a single node) 
is generated by clumping together the oscillating units of the previous $\mathbb{S}(1,K)$ graph (a single $K$-clique).

\begin{figure}
    \centering
    \includegraphics[width=0.35\textwidth]{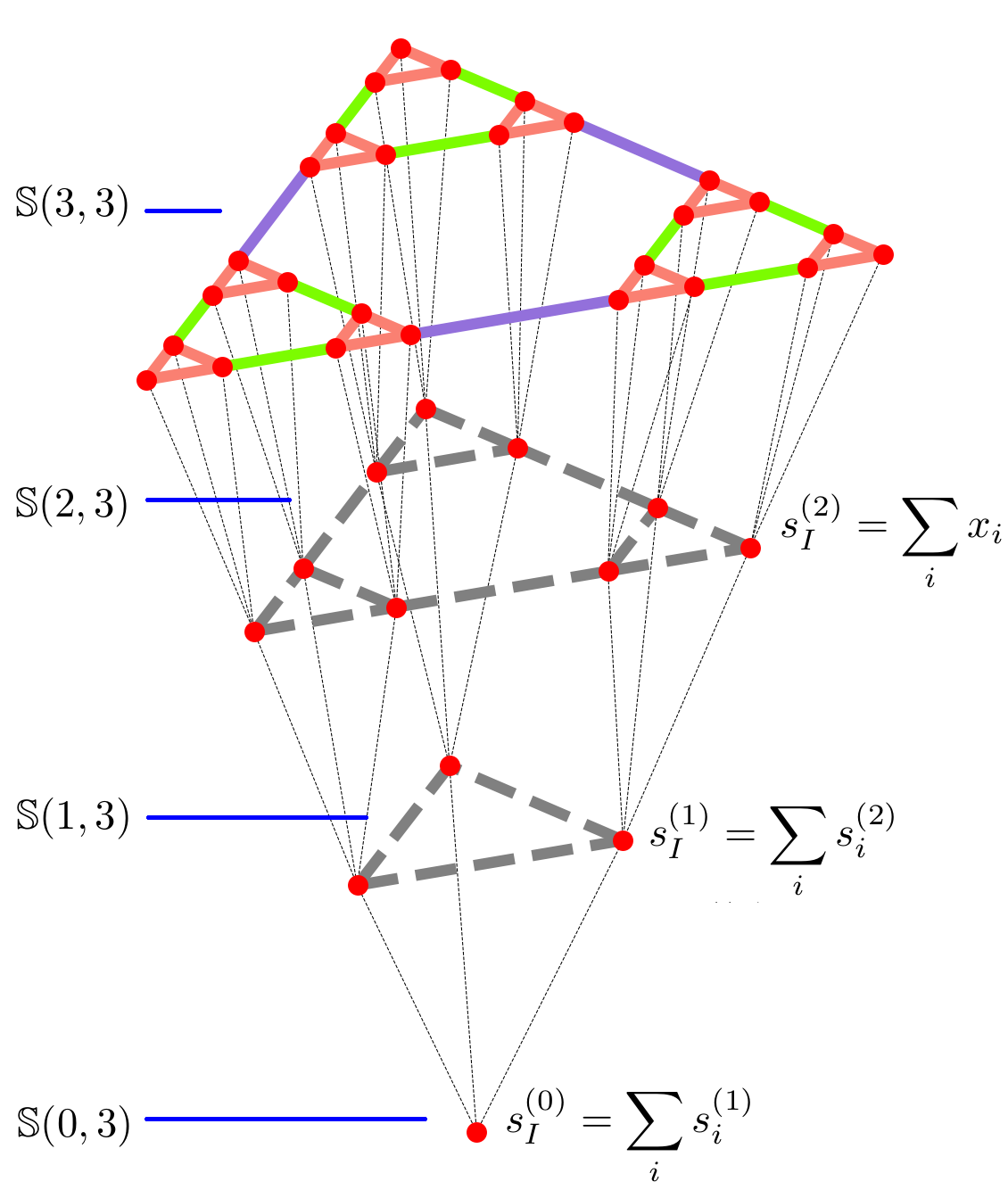}
    \caption{Illustration of the recursive coarse-graining analysis procedure. Starting from the base network of the Sierpinski graph topology $\mathbb{S}(3,3)$, nodes from each triangle or $3$-clique are grouped together into single nodes. The output $s_I$ of each new node  is the sum the of outputs of the $3$-clique oscillators forming the group ($s^{(2)}_I=\sum_{i} x_i$). The resulting nodes are connected following a Sierpinski graph $\mathbb{S}(2,3)$ layout, forming a coarse-grained system one level lower than the previous one. In the next step, nodes belonging to $3$-cliques are grouped by  adding their outputs, forming a coarse-grained network $\mathbb{S}(1,3)$,  with output of each $I$ node given by $s^{(1)}_I=\sum_{i} s^{(2)}_i$. Following the same procedure,in the final step the three nodes are collapsed into a network of one node only, $\mathbb{S}(0,3$) with output ($s^{(0)}_I=\sum_{i} s^{(1)}_i$). 
    The color code is the same as in Fig.~\ref{fig:1a}, with pink, green and purple representing links within each unit, between units, and between units of units, respectively.}
    \label{fig:1b}
\end{figure}

In the method described above, for a given step of the coarsening process, each of its units is generated from a clique of units in the preceding network by clumping them together. In general, for a coarse-grained network $\mathbb{S}(\nu-1,K)$ with $\nu-1$ levels, the output $s^{(\nu-1)}_I$ of an oscillating unit $I$, is given by the sum of the outputs $s^{(\nu)}_i$ of the oscillating units belonging to the parent network (with $\nu$ levels) from which this network was coarse-grained, thus
\begin{equation}
s^{(\nu-1)}_I = \sum_{i \in \,\text{a clique in $\mathbb{S}(\nu,K)$ }} s^{(\nu)}_i,
\end{equation}
where the sum is taken over the $K$ units in the corresponding clique of the parent network.  An example of this recursive procedure is shown in Fig.~\ref{fig:1b}, where an initial network $\mathbb{S}(3,3)$ is coarse-grained into a single unit ($\mathbb{S}(0,3)$) in three coarse-graining steps. 
In the figure the process of grouping together outputs from one network into a new output unit in the child network is illustrated. 
In this case, the three oscillators (units) in the rightmost clique of the base network $\mathbb{S}(3,3)$ are grouped together to generate the output $s^{(2)}_I$ in the singly coarse-grained 
network $\mathbb{S}(2,3)$. Following the same process, the rightmost clique of the second network is grouped together to generate the output $s^{(1)}_I$ in the twice coarse-grained 
network $\mathbb{S}(1,3)$, and finally the three last outputs in the network $\mathbb{S}(1,3)$ are collapsed into a single 
unit $\mathbb{S}(0,3)$.

\begin{figure}
    \centering
    \includegraphics[width=0.49\textwidth]{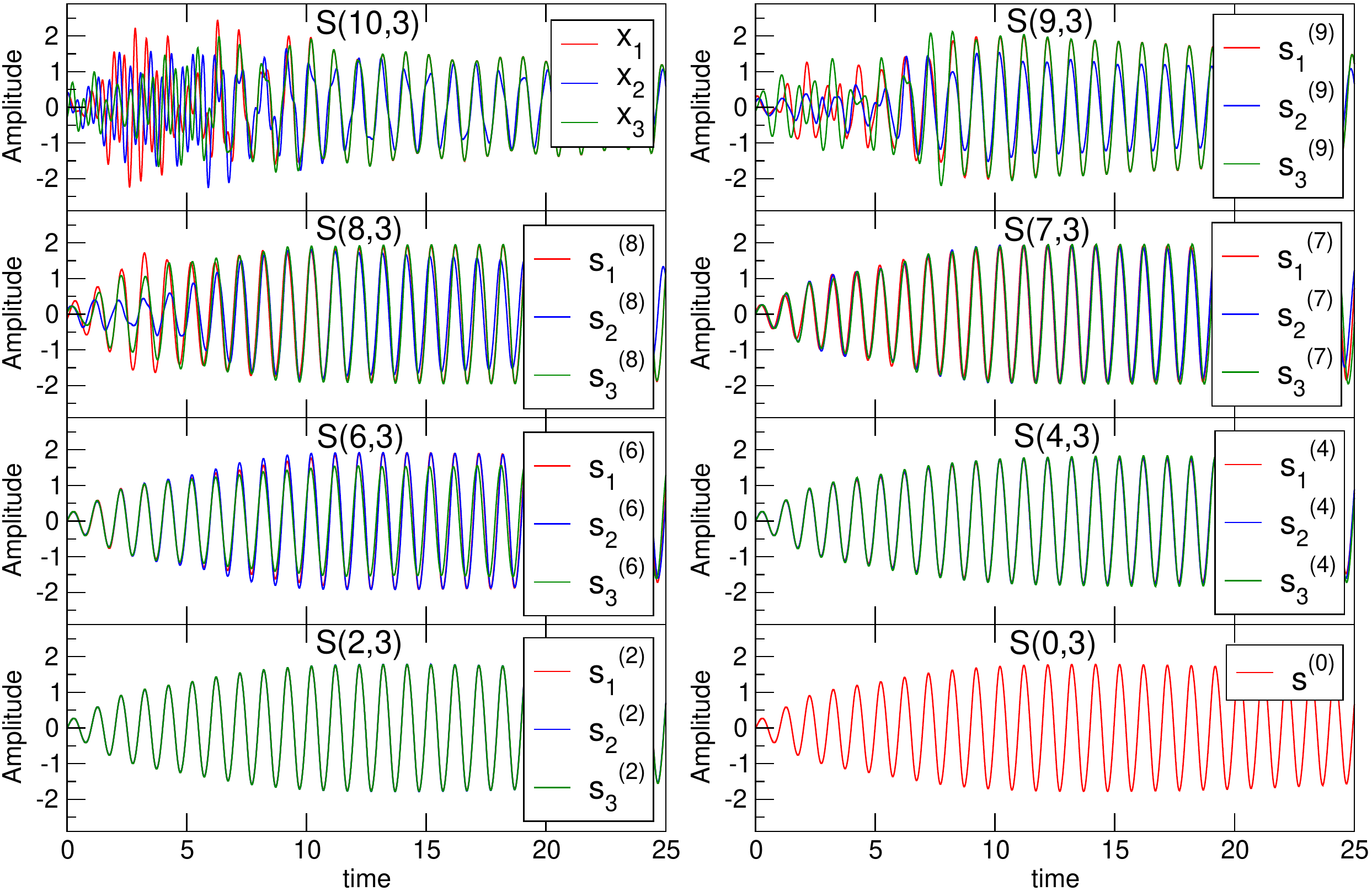}
    \caption{Time series of the measured signal at specific locations on the networks of different levels obtained during the coarse-graining process. In the top-left panel the amplitude $x$ of three different oscillators in a base network $\mathbb{S}(10,3)$ is shown. The signal of three different outputs in the coarse-grained networks $\mathbb{S}(9,3)$, $\mathbb{S}(8,3)$, $\mathbb{S}(7,3)$, $\mathbb{S}(6,3)$, $\mathbb{S}(4,3)$, $\mathbb{S}(2,3)$, and $\mathbb{S}(0,3)$, derived from the base network are shown in sequential order. For all these networks, 
    the three nodes chosen for visualization belonged to the first small triangle (from the left) located at one of the three corners of the corresponding network.
    }
    \label{fig:3}
\end{figure}

We study initially the temporal evolution of a system of hierarchically coupled oscillators connected following a Sierpinski graph $\mathbb{S}(10,3)$, solving numerically the system of equations~\ref{eq:2a}--\ref{eq:2b}. Initially the fundamental frequency of every oscillator in the system was fixed to $\omega_0$,  
the time step of integration with RK4 was set to $\delta t = 0.005 \omega_0^{-1}$, and the total time of each simulatios was $t_f=25$ time units. In the top-left panel of Fig.~\ref{fig:3} we show the temporal evolution (in units of $1/\omega_0$) 
of three oscillators located in one of the corner cliques (with a node with two links and the other two nodes with three links) of the 
network $\mathbb{S}(10,3)$.  We have tuned the parameters of our model system to be $a=0.4$, $q=0.7$, and $v=1$, in order to ensure that the final frequency stabilizes to one half of the fundamental frequency $\omega_0$, which resembles a system with two distinguishable rhythms like the one discussed in the Introduction showing circasemidian and circadian rhythms.  

Once we had numerically solved the dynamical equations, we applied the coarse-graining process to the system and observed the temporal evolution of the outputs over the recursive process. 
In the eight panels of Fig.~\ref{fig:3} we show the results for the three nodes of the coarse-grained corner clique in the $\mathbb{S}(9,3)$, $\mathbb{S}(8,3)$, $\mathbb{S}(7,3)$, $\mathbb{S}(6,3)$, $\mathbb{S}(4,3)$, $\mathbb{S}(2,3)$, and $\mathbb{S}(0,3)$ networks, respectively. 
It can be seen that the oscillators reach a synchronized state in about 10 time units (measured in units of $1/\omega_0$)  and that the coarse-grained network nodes at different levels of hierarchy oscillate with a frequency of about one half
of the fundamental frequency $\omega_0$ of the base oscillators.

The expected dynamics of the coupled system described in the Model section is that the fractal nature of the network connecting the oscillators induces the system to shift toward a synchronized state, with its frequency of oscillation tending towards an asymptotic value. This depends on the fundamental baseline frequency $\omega_0$ of each oscillator in the network and on the bifurcation parameter $v$ 
of the Van der Pol oscillator as well as on the network parameters, $a$, $q$, $K$ and $L$, i.e. the base level interaction strength, its decay factor, coordination number, and number of levels, respectively, but not on the number of nodes $N$ in the network. From Fig.~\ref{fig:3} it can be seen that the chosen oscillators of the base network 
approach a synchronized state  after a transient period. Here the frequency of oscillation settles to $\approx 0.55  \omega_0$, which is about one half of the fundamental frequency of the Van der Pol oscillator ($\omega_0$). Similarly, applying the coarse-graining procedure $L$ times we end up with the final single node ``network'' that oscillates with an asymptotic frequency of about 
one half of the fundamental frequency. 

\begin{figure}
    \centering
    \includegraphics[width=0.4\textwidth]{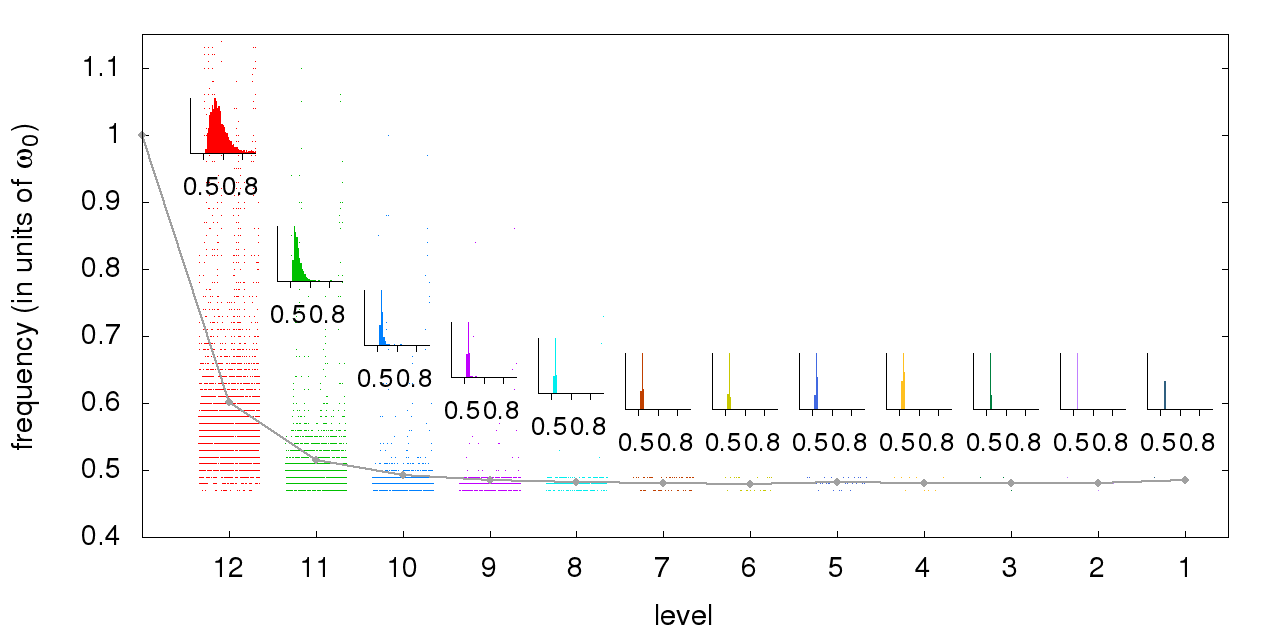}
    \includegraphics[width=0.4\textwidth]{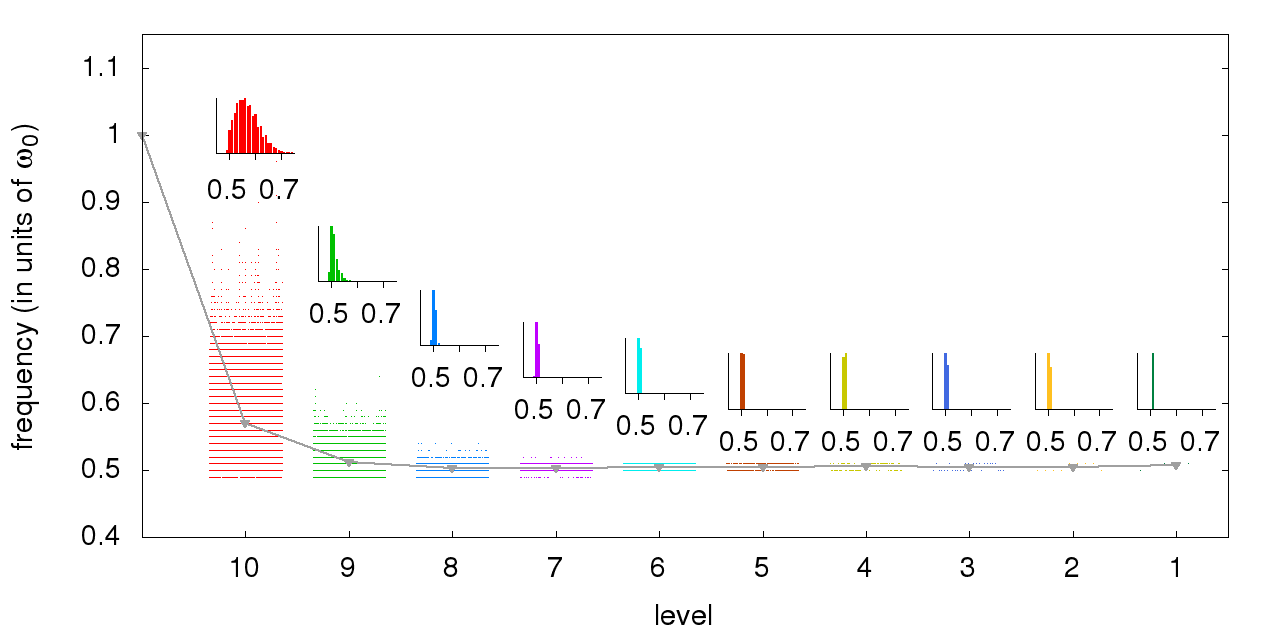}
    \includegraphics[width=0.4\textwidth]{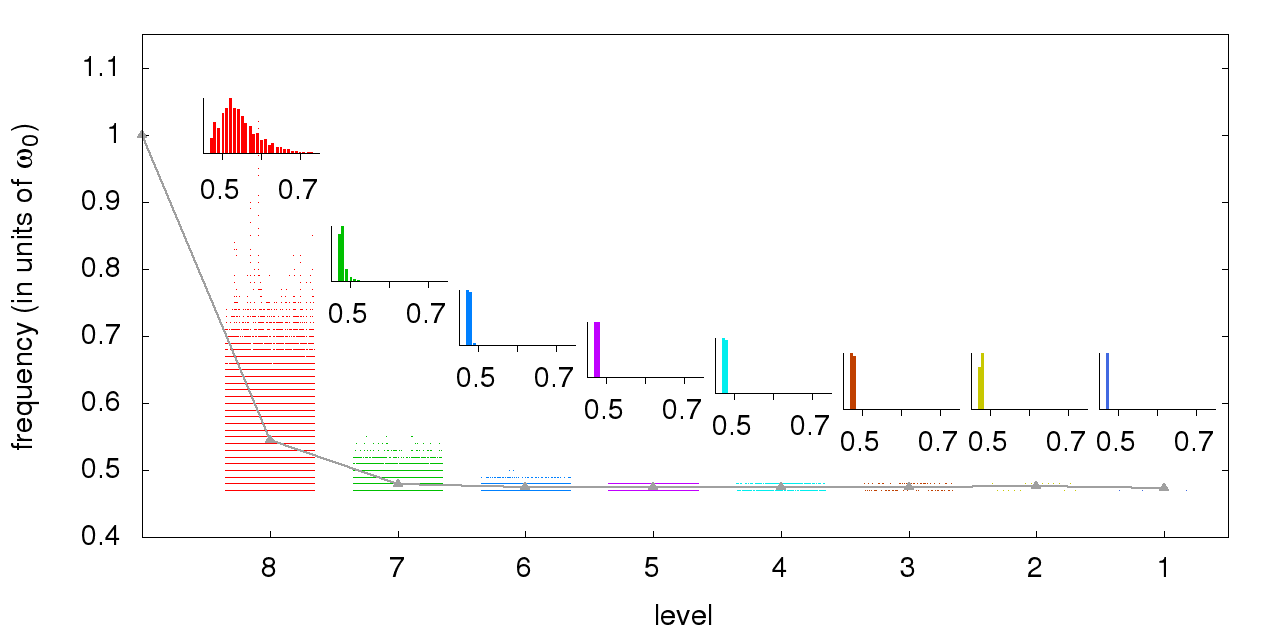}
    \includegraphics[width=0.4\textwidth]{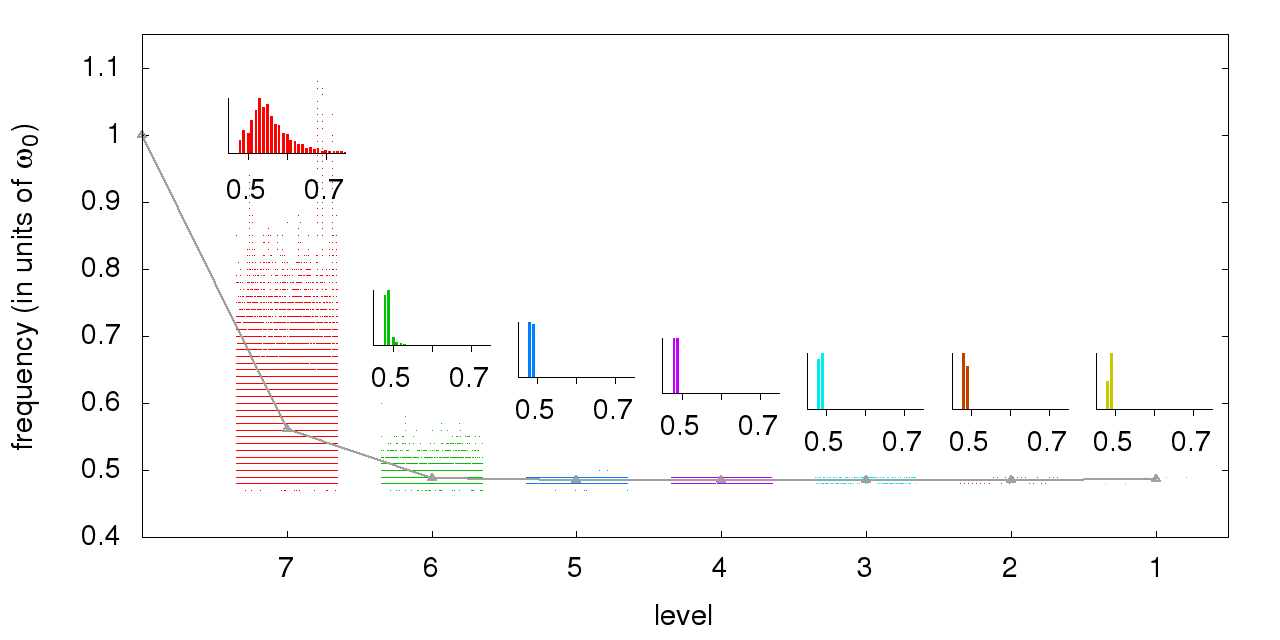}
    \includegraphics[width=0.4\textwidth]{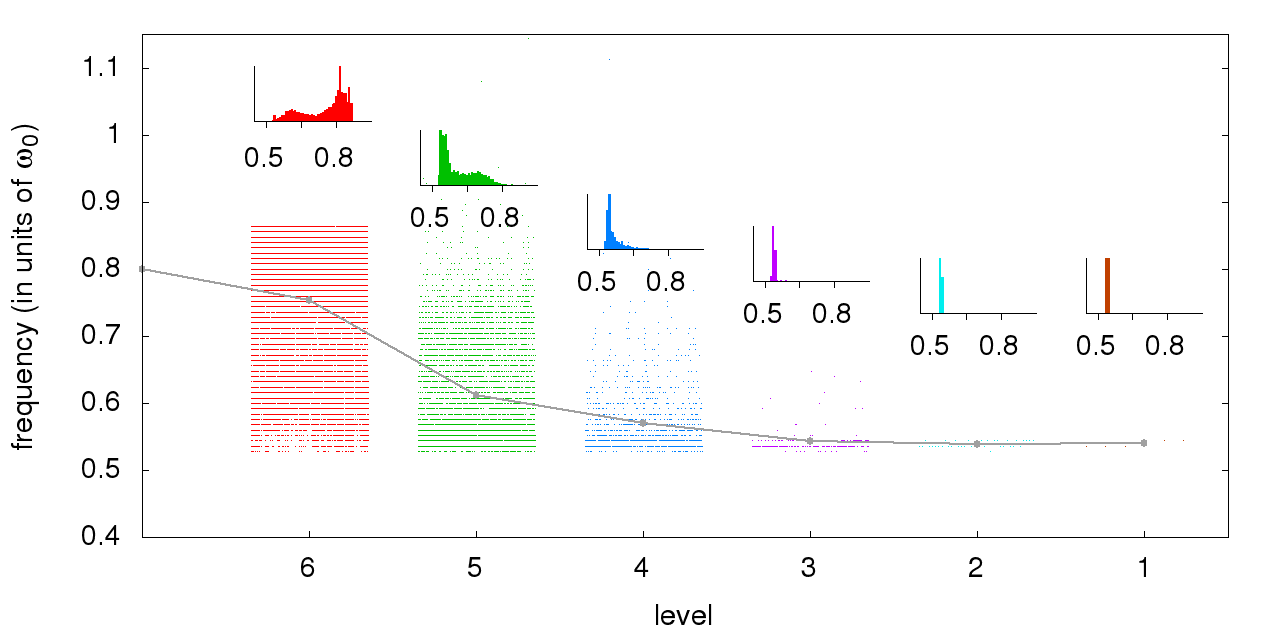}
    \caption{The mean of the observed frequency $\bar{\omega}$ (in units of $\omega_0$) of the system as a function of the number of levels $\nu$ in the coarse-grained network (gray line). The distribution of measured frequencies of the outputs for each level is shown in the small histogram above the corresponding mean value, with horizontal axis representing frequency values (in units of $\omega_0$). In addition, for each level, the frequency of each output is plotted (dots around the mean value), arranged horizontally in strict numbering order to ease the visualization. The network parameters (top) are $K=2$, $L=12$, 4096 nodes, $a=0.55$ and $q=0.8$. (middle-top) $K=3$, $L=10$, 59049 nodes, $a=0.4$ and $q=0.7$. (middle) $K=4$, $L=8$, 65536 nodes, $a=0.3$ and $q=0.7$. (middle-bottom) $K=5$, $L=7$, 16384 nodes, $a=0.2$ and $q=0.8$. (bottom) $K=6$, $L=6$, 46656 nodes, $a=0.14$ and $q=0.7$. The asymptotic frequency, for 
    the last step of the coarse-graining process, i.e. the coarse-grained network with just one level, is nearly half of the fundamental frequency $\omega_0$ for 
    all the topologies evaluated.}
    \label{fig:4}
\end{figure}

To explore the synchronization process along the coarse-graining procedure, we have numerically integrated the dynamical equations describing the system for the following five 
topologically different hierarchically coupled systems: $\mathbb{S}(2,12)$, $\mathbb{S}(3,10)$, $\mathbb{S}(4,8)$, $\mathbb{S}(5,7)$, $\mathbb{S}(6,6)$ with 4096, 59049, 65536, 16384 and 46656 nodes, respectively. For all these cases, we have set the bifurcation parameter $v=1$ and the RK4 time integration step size $dt=0.01 \omega_0^{-1}$, where $\omega_0$ is the fundamental frequency of all the uncoupled oscillators. 
In order to make the system reach an asymptotic frequency of about one half of the fundamental frequency ($\omega_0$), we need to tune the base level interaction strength $a$ and its decay parameter $q$ for each topology of the hierarchical network, thus for $K=$ 2, 3, 4, 5, and 6, we have set the strength $a$ to 0.55, 0.4, 0.3, 0.2 and 0.14 and $q$ to 0.8, 0.7, 0.7, 0.8 and 0.7, respectively. All the calculations were done with random initial conditions for $x_i(0)=0.5\zeta$ and $y_i(0)=0.1+0.01\zeta$, with $\zeta$ having a flat distribution in the interval $(-1,1)$. 
For each network topology, we calculate the mean frequency $\bar{\omega}$ of oscillation at each coarse-graining level $\nu$, as the average of the frequencies over the set of oscillators/units in the network. The results are shown in Fig.~\ref{fig:4} and it can be seen that in all cases, after the coarsening process is applied, the (asymptotic) mean frequency is about one half of the initial fundamental frequency ($\omega_0$). For the hierarchical network topology with $K \leq $3, we observe that the mean frequency keeps reducing to level 4, after which it is stabilized to 1/2 of the original fundamental frequency for all the remaining hierarchy levels. For other network topologies ($K >$3) the frequency reaches its asymptotic value faster, in other words after the second level of the coarsening process. 

To get a descriptor of the characteristic frequency to be found in an oscillator/unit belonging to a coarse grained network with a certain number number of levels. For this, we calculate the mean frequency $\bar{\omega}(\nu)$ as the average of the individual frequencies $\omega^{(\nu)}_i$ over the set of all oscillators/units of the network with $\nu$ levels, given by
$$\bar{\omega}(\nu)=\frac{1}{K^\nu}\sum_{i \in \mathbb{S}(\nu,K)} \omega^{(\nu)}_i,$$    
with $K$ the coordination number of the network. The frequency $\omega^{(\nu)}_i=2\pi/T_i$ of output $i$ was measured by calculating the average time period $T_i$ between pairs of consecutive peaks, and only the second half of the time series was used when calculating the average to exclude the initial transients.
Nevertheless, the oscillators located in different regions of the network have different frequencies, such that they are distributed around the mean, with the observed shape and width of the frequency distribution depending on the hierarchy level parameter $\nu$. 
At the base level network, $\nu=L$,  
the frequencies are distributed within the interval $[0.5\omega_0,0.9\omega_0]$ for all the network topologies, as depicted by the leftmost frequency bands (in red) in all the panels of Fig.~\ref{fig:4}. However, as the system is coarse-grained, the width of the distribution is continuously reduced, and after some steps of the coarse-graining process the distribution has collapsed into a single value, as seen in all the panels by the frequency bands becoming narrower when the number of levels is decreased from $\nu=L$ to $\nu=1$. 
The parameter $K$ also influences the shrinking process of the width of the distribution of oscillation frequencies. For the coordination number of the hierarchical networks, $K=$ 2, 3, 4 and 5, the width of the distribution seems to decrease similarly, quite rapidly, and uniformly in the consecutive coarse-graining steps, in such a way that in the third step of the coarse-grained process ($\nu=L-3$) the distribution of frequencies has shrunk to almost a single-value. However, in the case $K=6$ (hexagons) a slower shrinking in the frequency distribution width can be seen for the first three steps, $\nu=L, L-1, L-2$ levels, followed by a more rapid decrease from $\nu=L-3$ to an almost single-valued frequency.

In order to measure the synchronization of the outputs at different steps of the coarse-graining process, for each hierarchical network we calculate the index of synchronization $R$ occurring between $K$ outputs which would constitute a clique in this network~\cite{biswas2014peak}. For this we calculate the instantaneous phase $\theta_j(\nu,t)$ of the output $s_j$ in the level $\nu$ at time $t$ as 

\begin{equation*}   
\theta_j(\nu,t)=\tan^{-1}\frac{\dot{s}_j(\nu,t)}{s_j(\nu,t)},
\end{equation*}
with $\dot{s_j}(\nu,t)$ being the time derivative of the output $s_j(\nu,t)$. Defining the phase difference $\Delta \theta_{jk}$ between two connected outputs $s_j$, $s_k$  as $\Delta \theta_{jk}(\nu,t)=\theta_j(\nu,t)-\theta_k(\nu,t)$, we calculate the index of synchronization $R(\nu)$ between all the outputs in a clique at the level $\nu$ as 
\begin{equation}
R(\nu)=\left | \frac{1}{t_f}\sum_{t=0}^{t_f-1}  \frac{2}{K(K{-}1)} \sum_{j=1}^{K-1} \sum_{k=j+1}^{K} e^{i\Delta\theta_{jk}(\nu,t)}  \right | ,
\label{eq:3} 
\end{equation}
where the double summation is taken over all pairs of outputs in the clique,  $t_f$ is the total time of observation and $i$$=$$\sqrt{-1}$.


 
\begin{figure}[!ht]
    \centering
  \includegraphics[width=0.45\textwidth]{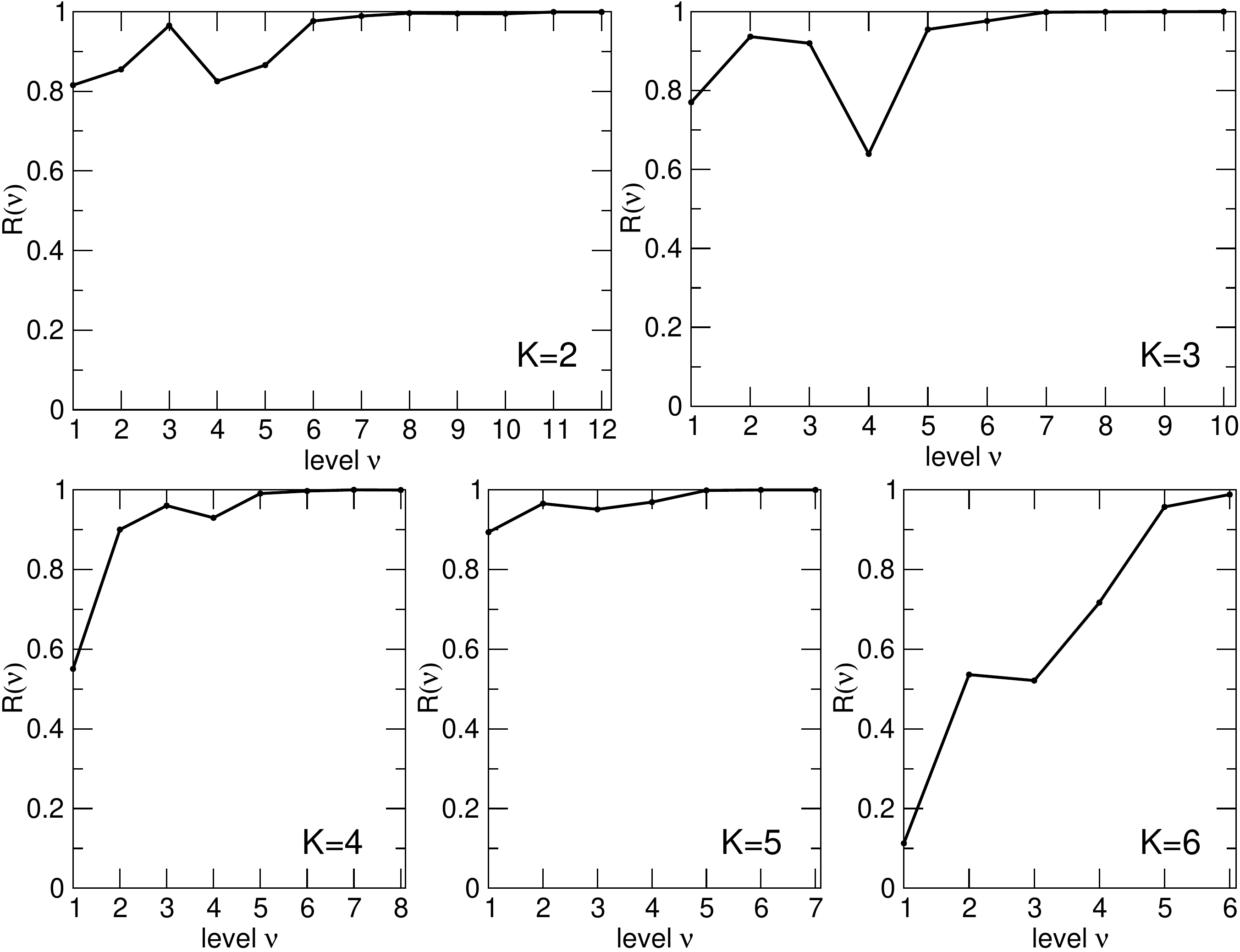}
    \caption{The index of synchronization $R(\nu)$ as a function of the level $\nu$, for hierarchical networks with different values of $K$.}
    \label{fig:5}
\end{figure}

In Fig.~\ref{fig:5} we plot the index of synchronization $R (\nu)$ as a function of the  level of hierarchy or coarse-graining, $\nu$, for the five different topologies with $K=$ 2, 3, 4, 5, and 6. The results indicate that for all these topologies, the oscillators/oscillating units inside each clique progressively reach a synchronized state when the level ($\nu$) of coarse-graining process is high enough, wherein the units are in phase.


\begin{figure}
    \centering
    \includegraphics[width=0.45\textwidth]{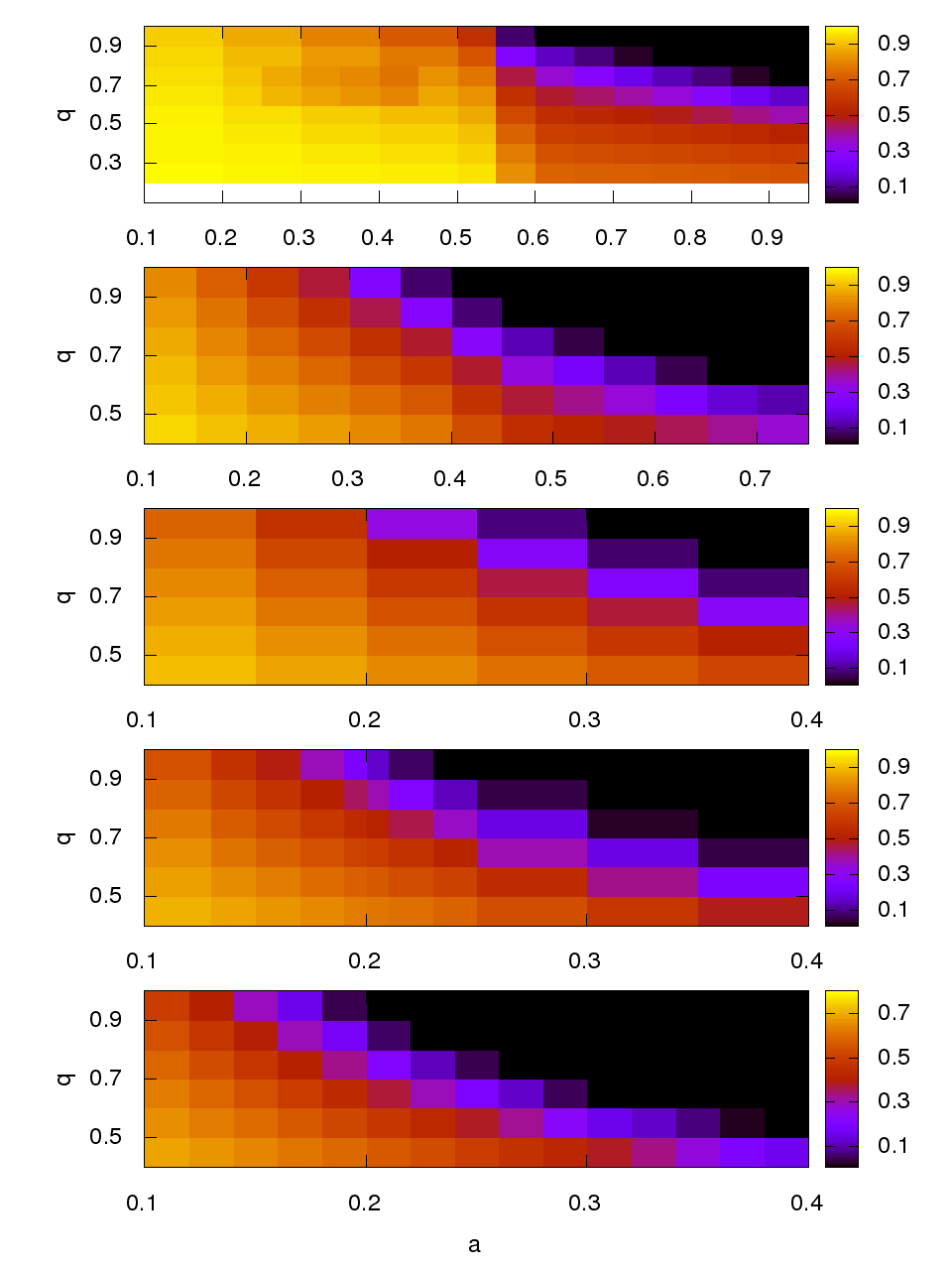}
\caption{Asymptotic frequency $\bar{\omega}$ over the parameter space $(a,q)$. Frequency is in units of the fundamental frequency $w_0$ of a Van Der Pol oscillator. (top) 2-clique, with 9 levels and 512 nodes. (middle-top) 3-clique, with 9 levels and 19683 nodes. (middle) 4-clique, with 7 levels and 16384 nodes. (middle-bottom) 5-clique, with 7 levels and 78125 nodes. (bottom) 6-clique, with 6 levels and 46656 nodes }
    \label{fig:6}
\end{figure}

In the results shown previously, the choice of the parameters, in particular the base interaction strength, $a$, and the decaying factor, $q$, was so that the system reached an asymptotic mean frequency $\bar{\omega}$ that is half of the fundamental 
frequency of a single Van der Pol oscillator to mimic the possible relation between the circadian and semi-circadian human cycle. Nevertheless, by a proper tuning of the parameters $a$ and $q$, 
the mean frequency $\bar{\omega}$ of the system can take any possible value in the interval  $\left[0,\omega_0\right]$, changing continuously as these parameters are varied. The extent and nature of the change in frequency at different levels of coarsening is explored next, by varying the parameters $a$ and $q$, and measuring the asymptotic mean frequency $\bar{\omega}$ that the system reaches in the final coarse-grained network.  The results for five hierarchical networks (with $K=$ 2, 3, 4, 5, and 6, and number of levels $12$, $9$, $7$, and $7$, respectively) are shown in Fig.~\ref{fig:6}. For all the cases, the base interaction strength $a$ was varied from $0.1$ to $0.9$, that is, from a loosely coupled to a tightly 
coupled system, whilst $q$, representing the amount of interaction decay between two consecutive levels of the hierarchy, was varied from $0.1$ to $0.9$.

The results in Fig.~\ref{fig:6} show a smooth transition between different frequencies as the parameters are changed. Inside the intervals explored, the asymptotic frequency can be tuned to any value between $0.4 \omega_0$ and $\omega_0$, and the transition is not abrupt nor discontinuous for any value of the parameters. It can be noticed that the size of the clique of the network influences the rate at which the asymptotic frequency falls as the parameters $a$ and $q$ are varied. In the case of the line and triangles ($K=$ 2, 3) the mean frequency $\bar{\omega}$ diminishes slowly, and larger values of $a$ and $q$ are required to force $\bar{\omega}$ to be zero, i.e. $a\approx 0.9$, $q\approx 0.8$ for $K=$2, and $a\approx 0.7$, $q\approx 0.7$ for $K=$3, respectively. On the other hand, for network topologies with larger cliques, i.e. $K\geq 4$, the decay of $\bar{\omega}$ is faster, with $K=$ 6 being the extreme case, where $\bar{\omega}$ tends to zero for values of $a$ and $q$ around 0.3 and 0.6, respectively.

\section{APPROXIMATE ANALYTIC SOLUTION}

In order to gain deeper insight into the properties of a hierarchical set of coupled oscillators, we consider an analytically tractable approximation. Take a single oscillator and couple $K$ oscillators with constant strength $r_{i}(0)=a$ (level $L=0$, see S(0,K) in Fig.~\ref{fig:1b}), to obtain Eqs.~\ref{eq:2a} and~\ref{eq:2b}. Add up the coordinates of the coupled oscillators and define new renormalised coordinates,

\begin{equation}
    X_1=\sum_{i=1}^{K}x_i,\;\;\;Y_1=\sum_{i=1}^{K}y_i.
\end{equation}  
This constitutes a new set of equations for the level $L=1$, 

\begin{subequations}
\begin{align}
    \dot{X}_1=&\omega_0Y_1 \label{eq:7a}\\
    \dot{Y}_1=&-\omega_0 \left(X_1-r_1(K-1)Y_1\right) +v\sum_{i=1}^K(1-x_i^2)y_i,\label{eq:7b}
\end{align}
\end{subequations}
where $r_1=qr_0=qa$. If we assume that the coordinates of the individual oscillators are uncorrelated, then the average of the products become the product of the averages ($\langle x^2y\rangle=\langle x^2\rangle  \langle y\rangle$) and one obtains a new renormalised Van der Pol oscillator:

\begin{equation}
\label{eq:8}
\ddot{X}_1=\omega_0^2\left[1-qr_0(K-1)\right] X_1+v(1-X_1^2)\dot{X}_1,
\end{equation}
whose 
frequency is $\omega_1^2=\omega_0^2\left[1-qr_0(K-1)\right]$. By repeating this procedure one can construct a set of $K^L$ oscillators vibrating with frequency $\omega_{L}^2\left[1-qr_{L-1}(K-1)\right]$, or

\begin{equation}
\label{eq:9}
\omega_{L}^2=\omega_0^2\prod_{i=0}^{L}\left[1-q^ia(K-1)\right] ,
\end{equation}
which in the limit of $L \rightarrow \infty$ has a finite non-zero value and is a well-known result, see~\cite{zwillinger2014table}. 

\begin{figure}[ht!]
    \centering
    \includegraphics[width=0.45\textwidth]{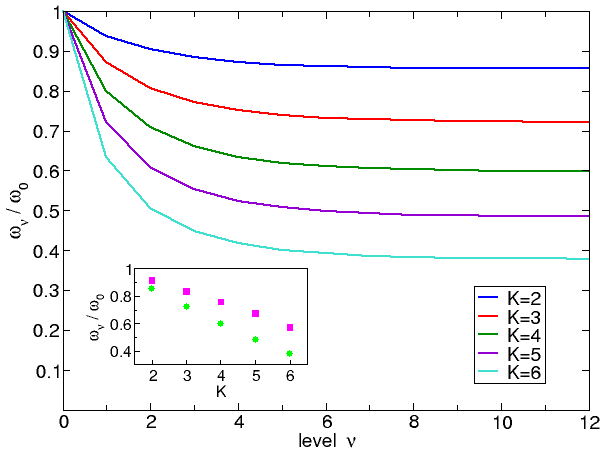}
\caption{ Frequency ${\omega}$ as a function of the number of steps $\nu$ of the coarse-graining process, derived from Eq.~\ref{eq:9} for $K=2,...,6$ and using the parameters  $(a=0.2,q=0.70)$. Frequency is in units of the fundamental frequency $w_0$ of a single Van Der Pol oscillator. Inset shows the comparison with the full model. Green circles refer to the analytic approximation of the frequency and magenta squares are the numerically computed frequencies of the full model.}
    \label{fig:7}
\end{figure}

In Fig.~\ref{fig:7} we show the behavior of the frequency as a function of the level of renormalisation for the cases $K=2,...,6$. Here one finds that the frequencies obtained from the numerical solution are systematically higher than those obtained by the analytical solutions, as depicted in the inset of Fig.~\ref{fig:7}. 
This is due to the fact that correlations present in the dynamical calculations tend to prevent the frequency from attaining its lower limit, which is only reached when there are no correlations.  One also notices that the differences between the predictions and the actual frequencies increase with $K$. This is to be expected, 
since the number of oscillators at 
the base network increases enormously when $K$ is large, thus leading to the omission of correlations in the system. 
Nevertheless, this approximation correctly captures the qualitative dependence of the frequency on the parameters $a$, $q$ and $k$. 

In Fig.~\ref{fig:4} one observes that the number of frequencies different from the asymptotic mean frequency decreases rapidly as one applies more steps of coarse-graining. Thus, the assumption in our approximate analytic solution that all the units at a given level oscillate with the same frequency is well supported when the number of coarse-graining steps is large, but to test this numerically for larger networks, like $K=6$ and $L=12$  having more than two billion oscillators, would take a prohibitively long time to compute. 
Nevertheless, we can conclude that our approximate analytical solution describes qualitatively the main effects found in the hierarchical networks, even if the dynamics are neglected.



\section{CONCLUDING REMARKS}
We have studied the dynamical properties of networks of Van der Pol oscillators with hierarchical couplings, the strength of which decrease as the number of hierarchical levels is increased. We characterized the collective behavior at every level of the hierarchy by coarse-graining the signals as outputs. 
From the outputs at each level of the hierarchy we have measured the mean frequency and the index of synchronization. Using these quantities we have demonstrated that the network can be tuned to synchronized states at different levels of hierarchy, with a characteristic mean frequency which is smaller than the fundamental frequency of an isolated Van Der Pol oscillator but different from zero. In addition, the asymptotic value of the  mean frequency can be stabilized to any desired value irrespective of the total number of oscillators in the system, depending exclusively on the coordination number $K$ of the network and on the coupling strength and its decay factor. 
These observations are 
supported by a mean-field-like approach, which captures qualitatively the dependence of the mean frequency on the coupling strength, its rate of decay with the hierarchy level, and the coordination number of the network.

Networks where coupling strengths show a broad distribution 
are common in social and biological systems ~\cite{monsivais2017seasonal,monsivais2017tracking,liu1997cellular}. 
Our study shows that in such networks a robust oscillation with a precise frequency can be obtained if the heterogeneity in coupling strengths is introduced using a hierarchy. Note, that previous studies on hierarchical coupling of oscillators were mostly concerned with topological hierarchies~\cite{ulonska2016chimera,PhysRevLett.96.114102}. 

Moreover, the phase locking phenomenon observed in our model is important, because it allows oscillations to be reset to any desired phase while the system is evolving dynamically. However, note that the phenomenon of synchronization in this system is different from the one that is 
observed in a system of weakly coupled Kuramoto  oscillators~\cite{strogatz2000kuramoto,boccaletti2002synchronization}. 

The present study is for an idealised system and future work will investigate how our results
are affected as we move to more realistic settings. For example, including disorder in hierarchical networks of oscillators, by varying the frequencies and phase of single oscillators. 
In addition, it would be interesting to explore more in depth the effects of the local topology in these hierarchical systems, by introducing local defects to break the clique symmetry or by using other fractal-like topologies, in order to test the robustness of the dynamics in less idealised topologies.
   
\section*{Acknowledgment}
DM, KB and KK  acknowledge support from the EU's H2020 Program under the scheme INFRAIA-1-2014-2015: Research Infrastructures", Grant agreement No. 654024 SoBigData: Social Mining and Big Data Ecosystem" (http://www.sobigdata.eu). RAB acknowledges financial support from Conacyt (Mexico) through project 283279. KK also acknowledges the Rutherford Foundation Visiting Fellowship at The Alan Turing Institute, UK.

\bibliography{ref.bib}

\begin{thebibliography}{31}%
\makeatletter
\providecommand \@ifxundefined [1]{%
 \@ifx{#1\undefined}
}%
\providecommand \@ifnum [1]{%
 \ifnum #1\expandafter \@firstoftwo
 \else \expandafter \@secondoftwo
 \fi
}%
\providecommand \@ifx [1]{%
 \ifx #1\expandafter \@firstoftwo
 \else \expandafter \@secondoftwo
 \fi
}%
\providecommand \natexlab [1]{#1}%
\providecommand \enquote  [1]{``#1''}%
\providecommand \bibnamefont  [1]{#1}%
\providecommand \bibfnamefont [1]{#1}%
\providecommand \citenamefont [1]{#1}%
\providecommand \href@noop [0]{\@secondoftwo}%
\providecommand \href [0]{\begingroup \@sanitize@url \@href}%
\providecommand \@href[1]{\@@startlink{#1}\@@href}%
\providecommand \@@href[1]{\endgroup#1\@@endlink}%
\providecommand \@sanitize@url [0]{\catcode `\\12\catcode `\$12\catcode
  `\&12\catcode `\#12\catcode `\^12\catcode `\_12\catcode `\%12\relax}%
\providecommand \@@startlink[1]{}%
\providecommand \@@endlink[0]{}%
\providecommand \url  [0]{\begingroup\@sanitize@url \@url }%
\providecommand \@url [1]{\endgroup\@href {#1}{\urlprefix }}%
\providecommand \urlprefix  [0]{URL }%
\providecommand \Eprint [0]{\href }%
\providecommand \doibase [0]{http://dx.doi.org/}%
\providecommand \selectlanguage [0]{\@gobble}%
\providecommand \bibinfo  [0]{\@secondoftwo}%
\providecommand \bibfield  [0]{\@secondoftwo}%
\providecommand \translation [1]{[#1]}%
\providecommand \BibitemOpen [0]{}%
\providecommand \bibitemStop [0]{}%
\providecommand \bibitemNoStop [0]{.\EOS\space}%
\providecommand \EOS [0]{\spacefactor3000\relax}%
\providecommand \BibitemShut  [1]{\csname bibitem#1\endcsname}%
\let\auto@bib@innerbib\@empty
\bibitem [{\citenamefont {Osipov}\ \emph {et~al.}(2007)\citenamefont {Osipov},
  \citenamefont {Kurths},\ and\ \citenamefont
  {Zhou}}]{osipov2007synchronization}%
  \BibitemOpen
  \bibfield  {author} {\bibinfo {author} {\bibfnamefont {G.~V.}\ \bibnamefont
  {Osipov}}, \bibinfo {author} {\bibfnamefont {J.}~\bibnamefont {Kurths}}, \
  and\ \bibinfo {author} {\bibfnamefont {C.}~\bibnamefont {Zhou}},\ }\href@noop
  {} {\emph {\bibinfo {title} {Synchronization in Oscillatory Networks}}}\
  (\bibinfo  {publisher} {Springer Science \& Business Media},\ \bibinfo {year}
  {2007})\BibitemShut {NoStop}%
\bibitem [{\citenamefont {Boccaletti}\ \emph {et~al.}(2002)\citenamefont
  {Boccaletti}, \citenamefont {Kurths}, \citenamefont {Osipov}, \citenamefont
  {Valladares},\ and\ \citenamefont {Zhou}}]{boccaletti2002synchronization}%
  \BibitemOpen
  \bibfield  {author} {\bibinfo {author} {\bibfnamefont {S.}~\bibnamefont
  {Boccaletti}}, \bibinfo {author} {\bibfnamefont {J.}~\bibnamefont {Kurths}},
  \bibinfo {author} {\bibfnamefont {G.}~\bibnamefont {Osipov}}, \bibinfo
  {author} {\bibfnamefont {D.}~\bibnamefont {Valladares}}, \ and\ \bibinfo
  {author} {\bibfnamefont {C.}~\bibnamefont {Zhou}},\ }\href@noop {} {\bibfield
   {journal} {\bibinfo  {journal} {Physics Reports}\ }\textbf {\bibinfo
  {volume} {366}},\ \bibinfo {pages} {1} (\bibinfo {year} {2002})}\BibitemShut
  {NoStop}%
\bibitem [{\citenamefont {Abrams}\ and\ \citenamefont
  {Strogatz}(2004)}]{abrams2004chimera}%
  \BibitemOpen
  \bibfield  {author} {\bibinfo {author} {\bibfnamefont {D.~M.}\ \bibnamefont
  {Abrams}}\ and\ \bibinfo {author} {\bibfnamefont {S.~H.}\ \bibnamefont
  {Strogatz}},\ }\href@noop {} {\bibfield  {journal} {\bibinfo  {journal}
  {Physical Review Letters}\ }\textbf {\bibinfo {volume} {93}},\ \bibinfo
  {pages} {174102} (\bibinfo {year} {2004})}\BibitemShut {NoStop}%
\bibitem [{\citenamefont {Strogatz}(2000)}]{strogatz2000kuramoto}%
  \BibitemOpen
  \bibfield  {author} {\bibinfo {author} {\bibfnamefont {S.~H.}\ \bibnamefont
  {Strogatz}},\ }\href@noop {} {\bibfield  {journal} {\bibinfo  {journal}
  {Physica D: Nonlinear Phenomena}\ }\textbf {\bibinfo {volume} {143}},\
  \bibinfo {pages} {1} (\bibinfo {year} {2000})}\BibitemShut {NoStop}%
\bibitem [{\citenamefont {Arenas}\ \emph {et~al.}(2008)\citenamefont {Arenas},
  \citenamefont {D{\'\i}az-Guilera}, \citenamefont {Kurths}, \citenamefont
  {Moreno},\ and\ \citenamefont {Zhou}}]{arenas2008synchronization}%
  \BibitemOpen
  \bibfield  {author} {\bibinfo {author} {\bibfnamefont {A.}~\bibnamefont
  {Arenas}}, \bibinfo {author} {\bibfnamefont {A.}~\bibnamefont
  {D{\'\i}az-Guilera}}, \bibinfo {author} {\bibfnamefont {J.}~\bibnamefont
  {Kurths}}, \bibinfo {author} {\bibfnamefont {Y.}~\bibnamefont {Moreno}}, \
  and\ \bibinfo {author} {\bibfnamefont {C.}~\bibnamefont {Zhou}},\ }\href@noop
  {} {\bibfield  {journal} {\bibinfo  {journal} {Physics Reports}\ }\textbf
  {\bibinfo {volume} {469}},\ \bibinfo {pages} {93} (\bibinfo {year}
  {2008})}\BibitemShut {NoStop}%
\bibitem [{\citenamefont {Goldbeter}(1997)}]{goldbeter1997biochemical}%
  \BibitemOpen
  \bibfield  {author} {\bibinfo {author} {\bibfnamefont {A.}~\bibnamefont
  {Goldbeter}},\ }\href@noop {} {\emph {\bibinfo {title} {Biochemical
  Oscillations and Cellular Rhythms: the Molecular Bases of Periodic and
  Chaotic Behaviour}}}\ (\bibinfo  {publisher} {Cambridge University Press},\
  \bibinfo {year} {1997})\BibitemShut {NoStop}%
\bibitem [{\citenamefont {Glass}(2001)}]{glass2001synchronization}%
  \BibitemOpen
  \bibfield  {author} {\bibinfo {author} {\bibfnamefont {L.}~\bibnamefont
  {Glass}},\ }\href@noop {} {\bibfield  {journal} {\bibinfo  {journal}
  {Nature}\ }\textbf {\bibinfo {volume} {410}},\ \bibinfo {pages} {277}
  (\bibinfo {year} {2001})}\BibitemShut {NoStop}%
\bibitem [{\citenamefont {Forger}(2017)}]{forger2017biological}%
  \BibitemOpen
  \bibfield  {author} {\bibinfo {author} {\bibfnamefont {D.~B.}\ \bibnamefont
  {Forger}},\ }\href@noop {} {\emph {\bibinfo {title} {Biological clocks,
  rhythms, and oscillations: the theory of biological timekeeping}}}\ (\bibinfo
  {year} {2017})\BibitemShut {NoStop}%
\bibitem [{\citenamefont {Liu}\ \emph {et~al.}(1997)\citenamefont {Liu},
  \citenamefont {Weaver}, \citenamefont {Strogatz},\ and\ \citenamefont
  {Reppert}}]{liu1997cellular}%
  \BibitemOpen
  \bibfield  {author} {\bibinfo {author} {\bibfnamefont {C.}~\bibnamefont
  {Liu}}, \bibinfo {author} {\bibfnamefont {D.~R.}\ \bibnamefont {Weaver}},
  \bibinfo {author} {\bibfnamefont {S.~H.}\ \bibnamefont {Strogatz}}, \ and\
  \bibinfo {author} {\bibfnamefont {S.~M.}\ \bibnamefont {Reppert}},\
  }\href@noop {} {\bibfield  {journal} {\bibinfo  {journal} {Cell}\ }\textbf
  {\bibinfo {volume} {91}},\ \bibinfo {pages} {855} (\bibinfo {year}
  {1997})}\BibitemShut {NoStop}%
\bibitem [{\citenamefont {Garcia-Ojalvo}\ \emph {et~al.}(2004)\citenamefont
  {Garcia-Ojalvo}, \citenamefont {Elowitz},\ and\ \citenamefont
  {Strogatz}}]{garcia2004modeling}%
  \BibitemOpen
  \bibfield  {author} {\bibinfo {author} {\bibfnamefont {J.}~\bibnamefont
  {Garcia-Ojalvo}}, \bibinfo {author} {\bibfnamefont {M.~B.}\ \bibnamefont
  {Elowitz}}, \ and\ \bibinfo {author} {\bibfnamefont {S.~H.}\ \bibnamefont
  {Strogatz}},\ }\href@noop {} {\bibfield  {journal} {\bibinfo  {journal}
  {Proceedings of the National Academy of Sciences}\ }\textbf {\bibinfo
  {volume} {101}},\ \bibinfo {pages} {10955} (\bibinfo {year}
  {2004})}\BibitemShut {NoStop}%
\bibitem [{\citenamefont {Barrio}\ \emph {et~al.}(1997)\citenamefont {Barrio},
  \citenamefont {Zhang},\ and\ \citenamefont
  {Maini}}]{barrio1997hierarchically}%
  \BibitemOpen
  \bibfield  {author} {\bibinfo {author} {\bibfnamefont {R.~A.}\ \bibnamefont
  {Barrio}}, \bibinfo {author} {\bibfnamefont {L.}~\bibnamefont {Zhang}}, \
  and\ \bibinfo {author} {\bibfnamefont {P.~K.}\ \bibnamefont {Maini}},\
  }\href@noop {} {\bibfield  {journal} {\bibinfo  {journal} {Bulletin of
  Mathematical Biology}\ }\textbf {\bibinfo {volume} {59}},\ \bibinfo {pages}
  {517} (\bibinfo {year} {1997})}\BibitemShut {NoStop}%
\bibitem [{\citenamefont {Grandin}\ \emph {et~al.}(2006)\citenamefont
  {Grandin}, \citenamefont {Alloy},\ and\ \citenamefont
  {Abramson}}]{grandin2006social}%
  \BibitemOpen
  \bibfield  {author} {\bibinfo {author} {\bibfnamefont {L.~D.}\ \bibnamefont
  {Grandin}}, \bibinfo {author} {\bibfnamefont {L.~B.}\ \bibnamefont {Alloy}},
  \ and\ \bibinfo {author} {\bibfnamefont {L.~Y.}\ \bibnamefont {Abramson}},\
  }\href@noop {} {\bibfield  {journal} {\bibinfo  {journal} {Clinical
  Psychology Review}\ }\textbf {\bibinfo {volume} {26}},\ \bibinfo {pages}
  {679} (\bibinfo {year} {2006})}\BibitemShut {NoStop}%
\bibitem [{\citenamefont {Monsivais}\ \emph
  {et~al.}(2017{\natexlab{a}})\citenamefont {Monsivais}, \citenamefont
  {Bhattacharya}, \citenamefont {Ghosh}, \citenamefont {Dunbar},\ and\
  \citenamefont {Kaski}}]{monsivais2017seasonal}%
  \BibitemOpen
  \bibfield  {author} {\bibinfo {author} {\bibfnamefont {D.}~\bibnamefont
  {Monsivais}}, \bibinfo {author} {\bibfnamefont {K.}~\bibnamefont
  {Bhattacharya}}, \bibinfo {author} {\bibfnamefont {A.}~\bibnamefont {Ghosh}},
  \bibinfo {author} {\bibfnamefont {R.~I.}\ \bibnamefont {Dunbar}}, \ and\
  \bibinfo {author} {\bibfnamefont {K.}~\bibnamefont {Kaski}},\ }\href@noop {}
  {\bibfield  {journal} {\bibinfo  {journal} {Scientific Reports}\ }\textbf
  {\bibinfo {volume} {7}},\ \bibinfo {pages} {10717} (\bibinfo {year}
  {2017}{\natexlab{a}})}\BibitemShut {NoStop}%
\bibitem [{\citenamefont {Monsivais}\ \emph
  {et~al.}(2017{\natexlab{b}})\citenamefont {Monsivais}, \citenamefont {Ghosh},
  \citenamefont {Bhattacharya}, \citenamefont {Dunbar},\ and\ \citenamefont
  {Kaski}}]{monsivais2017tracking}%
  \BibitemOpen
  \bibfield  {author} {\bibinfo {author} {\bibfnamefont {D.}~\bibnamefont
  {Monsivais}}, \bibinfo {author} {\bibfnamefont {A.}~\bibnamefont {Ghosh}},
  \bibinfo {author} {\bibfnamefont {K.}~\bibnamefont {Bhattacharya}}, \bibinfo
  {author} {\bibfnamefont {R.~I.}\ \bibnamefont {Dunbar}}, \ and\ \bibinfo
  {author} {\bibfnamefont {K.}~\bibnamefont {Kaski}},\ }\href@noop {}
  {\bibfield  {journal} {\bibinfo  {journal} {PLoS Computational Biology}\
  }\textbf {\bibinfo {volume} {13}},\ \bibinfo {pages} {e1005824} (\bibinfo
  {year} {2017}{\natexlab{b}})}\BibitemShut {NoStop}%
\bibitem [{\citenamefont {Mistlberger}\ and\ \citenamefont
  {Skene}(2004)}]{mistlberger2004social}%
  \BibitemOpen
  \bibfield  {author} {\bibinfo {author} {\bibfnamefont {R.~E.}\ \bibnamefont
  {Mistlberger}}\ and\ \bibinfo {author} {\bibfnamefont {D.~J.}\ \bibnamefont
  {Skene}},\ }\href@noop {} {\bibfield  {journal} {\bibinfo  {journal}
  {Biological Reviews}\ }\textbf {\bibinfo {volume} {79}},\ \bibinfo {pages}
  {533} (\bibinfo {year} {2004})}\BibitemShut {NoStop}%
\bibitem [{\citenamefont {Bloch}\ \emph {et~al.}(2013)\citenamefont {Bloch},
  \citenamefont {Herzog}, \citenamefont {Levine},\ and\ \citenamefont
  {Schwartz}}]{bloch2013socially}%
  \BibitemOpen
  \bibfield  {author} {\bibinfo {author} {\bibfnamefont {G.}~\bibnamefont
  {Bloch}}, \bibinfo {author} {\bibfnamefont {E.~D.}\ \bibnamefont {Herzog}},
  \bibinfo {author} {\bibfnamefont {J.~D.}\ \bibnamefont {Levine}}, \ and\
  \bibinfo {author} {\bibfnamefont {W.~J.}\ \bibnamefont {Schwartz}},\
  }\href@noop {} {\bibfield  {journal} {\bibinfo  {journal} {Proceedings of the
  Royal Society B: Biological Sciences}\ }\textbf {\bibinfo {volume} {280}},\
  \bibinfo {pages} {20130035} (\bibinfo {year} {2013})}\BibitemShut {NoStop}%
\bibitem [{\citenamefont {Fuchikawa}\ \emph {et~al.}(2016)\citenamefont
  {Fuchikawa}, \citenamefont {Eban-Rothschild}, \citenamefont {Nagari},
  \citenamefont {Shemesh},\ and\ \citenamefont {Bloch}}]{fuchikawa2016potent}%
  \BibitemOpen
  \bibfield  {author} {\bibinfo {author} {\bibfnamefont {T.}~\bibnamefont
  {Fuchikawa}}, \bibinfo {author} {\bibfnamefont {A.}~\bibnamefont
  {Eban-Rothschild}}, \bibinfo {author} {\bibfnamefont {M.}~\bibnamefont
  {Nagari}}, \bibinfo {author} {\bibfnamefont {Y.}~\bibnamefont {Shemesh}}, \
  and\ \bibinfo {author} {\bibfnamefont {G.}~\bibnamefont {Bloch}},\
  }\href@noop {} {\bibfield  {journal} {\bibinfo  {journal} {Nature
  Communications}\ }\textbf {\bibinfo {volume} {7}},\ \bibinfo {pages} {11662}
  (\bibinfo {year} {2016})}\BibitemShut {NoStop}%
\bibitem [{\citenamefont {Pastor}\ \emph {et~al.}(1993)\citenamefont {Pastor},
  \citenamefont {P{\'e}rez-Garc{\'\i}a}, \citenamefont {Encinas},\ and\
  \citenamefont {Guerra}}]{pastor1993ordered}%
  \BibitemOpen
  \bibfield  {author} {\bibinfo {author} {\bibfnamefont {I.}~\bibnamefont
  {Pastor}}, \bibinfo {author} {\bibfnamefont {V.~M.}\ \bibnamefont
  {P{\'e}rez-Garc{\'\i}a}}, \bibinfo {author} {\bibfnamefont {F.}~\bibnamefont
  {Encinas}}, \ and\ \bibinfo {author} {\bibfnamefont {J.}~\bibnamefont
  {Guerra}},\ }\href@noop {} {\bibfield  {journal} {\bibinfo  {journal}
  {Physical Review E}\ }\textbf {\bibinfo {volume} {48}},\ \bibinfo {pages}
  {171} (\bibinfo {year} {1993})}\BibitemShut {NoStop}%
\bibitem [{\citenamefont {Ulonska}\ \emph {et~al.}(2016)\citenamefont
  {Ulonska}, \citenamefont {Omelchenko}, \citenamefont {Zakharova},\ and\
  \citenamefont {Sch{\"o}ll}}]{ulonska2016chimera}%
  \BibitemOpen
  \bibfield  {author} {\bibinfo {author} {\bibfnamefont {S.}~\bibnamefont
  {Ulonska}}, \bibinfo {author} {\bibfnamefont {I.}~\bibnamefont {Omelchenko}},
  \bibinfo {author} {\bibfnamefont {A.}~\bibnamefont {Zakharova}}, \ and\
  \bibinfo {author} {\bibfnamefont {E.}~\bibnamefont {Sch{\"o}ll}},\
  }\href@noop {} {\bibfield  {journal} {\bibinfo  {journal} {Chaos: An
  Interdisciplinary Journal of Nonlinear Science}\ }\textbf {\bibinfo {volume}
  {26}},\ \bibinfo {pages} {094825} (\bibinfo {year} {2016})}\BibitemShut
  {NoStop}%
\bibitem [{\citenamefont {D{\"o}rfler}\ and\ \citenamefont
  {Bullo}(2014)}]{dorfler2014synchronization}%
  \BibitemOpen
  \bibfield  {author} {\bibinfo {author} {\bibfnamefont {F.}~\bibnamefont
  {D{\"o}rfler}}\ and\ \bibinfo {author} {\bibfnamefont {F.}~\bibnamefont
  {Bullo}},\ }\href@noop {} {\bibfield  {journal} {\bibinfo  {journal}
  {Automatica}\ }\textbf {\bibinfo {volume} {50}},\ \bibinfo {pages} {1539}
  (\bibinfo {year} {2014})}\BibitemShut {NoStop}%
\bibitem [{\citenamefont {Perlikowski}\ \emph {et~al.}(2010)\citenamefont
  {Perlikowski}, \citenamefont {Stefanski},\ and\ \citenamefont
  {Kapitaniak}}]{perlikowski2010discontinuous}%
  \BibitemOpen
  \bibfield  {author} {\bibinfo {author} {\bibfnamefont {P.}~\bibnamefont
  {Perlikowski}}, \bibinfo {author} {\bibfnamefont {A.}~\bibnamefont
  {Stefanski}}, \ and\ \bibinfo {author} {\bibfnamefont {T.}~\bibnamefont
  {Kapitaniak}},\ }\href@noop {} {\bibfield  {journal} {\bibinfo  {journal}
  {International Journal of Non-Linear Mechanics}\ }\textbf {\bibinfo {volume}
  {45}},\ \bibinfo {pages} {895} (\bibinfo {year} {2010})}\BibitemShut
  {NoStop}%
\bibitem [{\citenamefont {Hizanidis}\ \emph {et~al.}(2015)\citenamefont
  {Hizanidis}, \citenamefont {Panagakou}, \citenamefont {Omelchenko},
  \citenamefont {Sch{\"o}ll}, \citenamefont {H{\"o}vel},\ and\ \citenamefont
  {Provata}}]{hizanidis2015chimera}%
  \BibitemOpen
  \bibfield  {author} {\bibinfo {author} {\bibfnamefont {J.}~\bibnamefont
  {Hizanidis}}, \bibinfo {author} {\bibfnamefont {E.}~\bibnamefont
  {Panagakou}}, \bibinfo {author} {\bibfnamefont {I.}~\bibnamefont
  {Omelchenko}}, \bibinfo {author} {\bibfnamefont {E.}~\bibnamefont
  {Sch{\"o}ll}}, \bibinfo {author} {\bibfnamefont {P.}~\bibnamefont
  {H{\"o}vel}}, \ and\ \bibinfo {author} {\bibfnamefont {A.}~\bibnamefont
  {Provata}},\ }\href@noop {} {\bibfield  {journal} {\bibinfo  {journal}
  {Physical Review E}\ }\textbf {\bibinfo {volume} {92}},\ \bibinfo {pages}
  {012915} (\bibinfo {year} {2015})}\BibitemShut {NoStop}%
\bibitem [{\citenamefont {Bera}\ \emph {et~al.}(2016)\citenamefont {Bera},
  \citenamefont {Ghosh},\ and\ \citenamefont {Banerjee}}]{bera2016imperfect}%
  \BibitemOpen
  \bibfield  {author} {\bibinfo {author} {\bibfnamefont {B.~K.}\ \bibnamefont
  {Bera}}, \bibinfo {author} {\bibfnamefont {D.}~\bibnamefont {Ghosh}}, \ and\
  \bibinfo {author} {\bibfnamefont {T.}~\bibnamefont {Banerjee}},\ }\href@noop
  {} {\bibfield  {journal} {\bibinfo  {journal} {Physical Review E}\ }\textbf
  {\bibinfo {volume} {94}},\ \bibinfo {pages} {012215} (\bibinfo {year}
  {2016})}\BibitemShut {NoStop}%
\bibitem [{\citenamefont {Krishnagopal}\ \emph {et~al.}(2017)\citenamefont
  {Krishnagopal}, \citenamefont {Lehnert}, \citenamefont {Poel}, \citenamefont
  {Zakharova},\ and\ \citenamefont
  {Sch{\"o}ll}}]{krishnagopal2017synchronization}%
  \BibitemOpen
  \bibfield  {author} {\bibinfo {author} {\bibfnamefont {S.}~\bibnamefont
  {Krishnagopal}}, \bibinfo {author} {\bibfnamefont {J.}~\bibnamefont
  {Lehnert}}, \bibinfo {author} {\bibfnamefont {W.}~\bibnamefont {Poel}},
  \bibinfo {author} {\bibfnamefont {A.}~\bibnamefont {Zakharova}}, \ and\
  \bibinfo {author} {\bibfnamefont {E.}~\bibnamefont {Sch{\"o}ll}},\
  }\href@noop {} {\bibfield  {journal} {\bibinfo  {journal} {Philosophical
  Transactions of the Royal Society A: Mathematical, Physical and Engineering
  Sciences}\ }\textbf {\bibinfo {volume} {375}},\ \bibinfo {pages} {20160216}
  (\bibinfo {year} {2017})}\BibitemShut {NoStop}%
\bibitem [{\citenamefont {Rakshit}\ \emph {et~al.}(2018)\citenamefont
  {Rakshit}, \citenamefont {Bera},\ and\ \citenamefont
  {Ghosh}}]{rakshit2018synchronization}%
  \BibitemOpen
  \bibfield  {author} {\bibinfo {author} {\bibfnamefont {S.}~\bibnamefont
  {Rakshit}}, \bibinfo {author} {\bibfnamefont {B.~K.}\ \bibnamefont {Bera}}, \
  and\ \bibinfo {author} {\bibfnamefont {D.}~\bibnamefont {Ghosh}},\
  }\href@noop {} {\bibfield  {journal} {\bibinfo  {journal} {Physical Review
  E}\ }\textbf {\bibinfo {volume} {98}},\ \bibinfo {pages} {032305} (\bibinfo
  {year} {2018})}\BibitemShut {NoStop}%
\bibitem [{\citenamefont {Van Der~Pol}(1927)}]{van1927vii}%
  \BibitemOpen
  \bibfield  {author} {\bibinfo {author} {\bibfnamefont {B.}~\bibnamefont {Van
  Der~Pol}},\ }\href@noop {} {\bibfield  {journal} {\bibinfo  {journal} {The
  London, Edinburgh, and Dublin Philosophical Magazine and Journal of Science}\
  }\textbf {\bibinfo {volume} {3}},\ \bibinfo {pages} {65} (\bibinfo {year}
  {1927})}\BibitemShut {NoStop}%
\bibitem [{Note1()}]{Note1}%
  \BibitemOpen
  \bibinfo {note} {The Van der Pol equation is a special case of the Rayleigh
  Differential equation}\BibitemShut {NoStop}%
\bibitem [{\citenamefont {Klav{\v{z}}ar}\ and\ \citenamefont
  {Milutinovi{\'c}}(1997)}]{klavvzar1997graphs}%
  \BibitemOpen
  \bibfield  {author} {\bibinfo {author} {\bibfnamefont {S.}~\bibnamefont
  {Klav{\v{z}}ar}}\ and\ \bibinfo {author} {\bibfnamefont {U.}~\bibnamefont
  {Milutinovi{\'c}}},\ }\href@noop {} {\bibfield  {journal} {\bibinfo
  {journal} {Czechoslovak Mathematical Journal}\ }\textbf {\bibinfo {volume}
  {47}},\ \bibinfo {pages} {95} (\bibinfo {year} {1997})}\BibitemShut {NoStop}%
\bibitem [{\citenamefont {Biswas}\ \emph {et~al.}(2014)\citenamefont {Biswas},
  \citenamefont {Khamaru},\ and\ \citenamefont {Majumdar}}]{biswas2014peak}%
  \BibitemOpen
  \bibfield  {author} {\bibinfo {author} {\bibfnamefont {R.}~\bibnamefont
  {Biswas}}, \bibinfo {author} {\bibfnamefont {K.}~\bibnamefont {Khamaru}}, \
  and\ \bibinfo {author} {\bibfnamefont {K.~K.}\ \bibnamefont {Majumdar}},\
  }\href@noop {} {\bibfield  {journal} {\bibinfo  {journal} {IEEE Transactions
  on Signal Processing}\ }\textbf {\bibinfo {volume} {62}},\ \bibinfo {pages}
  {4390} (\bibinfo {year} {2014})}\BibitemShut {NoStop}%
\bibitem [{\citenamefont {Zwillinger}(2014)}]{zwillinger2014table}%
  \BibitemOpen
  \bibfield  {author} {\bibinfo {author} {\bibfnamefont {D.}~\bibnamefont
  {Zwillinger}},\ }\href@noop {} {\emph {\bibinfo {title} {Table of integrals,
  series, and products}}}\ (\bibinfo  {publisher} {Elsevier},\ \bibinfo {year}
  {2014})\BibitemShut {NoStop}%
\bibitem [{\citenamefont {Arenas}\ \emph {et~al.}(2006)\citenamefont {Arenas},
  \citenamefont {D\'{\i}az-Guilera},\ and\ \citenamefont
  {P\'erez-Vicente}}]{PhysRevLett.96.114102}%
  \BibitemOpen
  \bibfield  {author} {\bibinfo {author} {\bibfnamefont {A.}~\bibnamefont
  {Arenas}}, \bibinfo {author} {\bibfnamefont {A.}~\bibnamefont
  {D\'{\i}az-Guilera}}, \ and\ \bibinfo {author} {\bibfnamefont {C.~J.}\
  \bibnamefont {P\'erez-Vicente}},\ }\href {\doibase
  10.1103/PhysRevLett.96.114102} {\bibfield  {journal} {\bibinfo  {journal}
  {Phys. Rev. Lett.}\ }\textbf {\bibinfo {volume} {96}},\ \bibinfo {pages}
  {114102} (\bibinfo {year} {2006})}\BibitemShut {NoStop}%
\end{thebibliography}%

\end{document}